\documentclass[aps,rmp,reprint,amssymb,longbibliography]{revtex4-1}

\usepackage{graphicx}	
\usepackage{array}
\usepackage{longtable}
\usepackage{amsmath}
\usepackage{url}
\usepackage{xcolor}

\newcolumntype{L}[1]{>{\raggedright\let\newline\\\arraybackslash\hspace{0pt}}m{#1}}

\newcommand{\nb}[1]{#1}
\newcommand{\ph}[1]{#1}

\begin{document}

\title{\textit{Colloquium:} Neutrino Detectors as Tools for Nuclear Security}

\author{Adam Bernstein}
\email{bernstein3@llnl.gov}

\author{Nathaniel Bowden}%
\email{nbowden@llnl.gov}
\affiliation{Nuclear and Chemical Sciences Division, Lawrence Livermore National Laboratory, Livermore, CA 94550}

\author{Bethany L. Goldblum}
\email{bethany@nuc.berkeley.edu}
\affiliation{Department of Nuclear Engineering, University of California, Berkeley, CA 94720}

\author{Patrick Huber}
\email{pahuber@vt.edu}
\affiliation{Center for Neutrino Physics, Virginia Tech, Blacksburg, VA 24061}

\author{Igor Jovanovic}
\email{ijov@umich.edu}
\affiliation{Department of Nuclear Engineering and Radiological Sciences, University of Michigan, Ann Arbor, MI 48109}

\author{John Mattingly}
\email{jkmattin@ncsu.edu}
\affiliation{Department of Nuclear Engineering, North Carolina State University, Raleigh, NC 27695}

\date{\today}
\begin{abstract}
For over 40 years, physicists have considered possible uses for
neutrino detectors in nuclear nonproliferation, arms control, and
fissile materials security. Neutrinos are an attractive fission
signature because they readily pass through matter. The same property
makes neutrinos challenging to detect in systems that would be
practical for nuclear security applications. This colloquium presents a
broad overview of several potential neutrino applications, including
the near-field monitoring of known reactors, far-field monitoring of
known or discovery of undeclared reactors, detection of reactor waste
streams, and detection of nuclear explosions. We conclude that recent
detector advances have made near-field monitoring
feasible. Farther-field reactor detection and waste stream detection
monitoring are possible in some cases with further research and
development. Very long-range reactor monitoring and nuclear explosion
detection do not appear feasible for the foreseeable future due to
considerable physical and/or practical constraints.
\end{abstract}


\maketitle

\tableofcontents{}

\section{Introduction}
\label{sec:intro}

The advent of nuclear weapons as the first practical application of
nuclear fission profoundly affected the dynamics of international
relations. The destructive potential of nuclear weapons rendered
conflicts in which they could be used potentially catastrophic, with
weapons effects far surpassing those of conventional armaments. The
effects of nuclear weapons cannot be constrained to the location where
they are used, because of the subsequent radioactive fallout and
potential multi-year effects on the global climate. While the United
States and the USSR avoided using nuclear weapons throughout the Cold
War, both took part in an arms race that, at its apex in 1986, resulted in a
stockpile of an estimated 63,000 warheads \cite{Kristensen2013}. During the Cold War and
afterwards, nuclear weapons proliferated, the production of special
nuclear materials continued, and nuclear knowledge spread across the
globe, even in states that did not have nuclear weapons -- creating
another major risk, nuclear terrorism. Today we are faced with nine
countries having a total of nearly 15,000 nuclear weapons and there
are additional countries at the verge of or actively seeking a nuclear
weapons capability \cite{Kristensen2017}.

Recognition of these unique challenges led to major international
efforts to curb the testing and use of nuclear technology for the
purpose of nuclear warfare and to bolster nuclear security. To gain
more coherence and legitimacy, these efforts have been articulated
through several international treaties -- most notably the Treaty on
the Non-proliferation of Nuclear Weapons (NPT), which came into force
in 1970. While the NPT provides an institutional and legal framework
to curb proliferation, it also requires the development and adoption
of effective technical measures for verification.

Applied antineutrino physics has the potential to provide novel
verification technologies, especially with regard to plutonium
production and diversion. First we give a brief summary of the
current safeguards framework. Next, we provide an overview of the current
state of knowledge and opportunities for future technical developments
in the area of antineutrino detection for nuclear security, with a
focus on four areas: monitoring of fissile material production,
discovery and exclusion of undeclared reactors, monitoring of spent
fuel and reprocessing waste, and confirmation of nuclear
explosions. For each of these applications we discuss the current technical
means of verification and highlight additional capabilities
offered by antineutrino detection.

\section{Current safeguards framework}
\label{sec:framework}

The Treaty on the Non-Proliferation of Nuclear Weapons~\cite{NPT} is
the central pillar of the international legal framework addressing the
security challenges arising from nuclear weapons. It has been in force
since 1970 and has 191 signatories, making it the most widely accepted
arms control and disarmament agreement to date.

The control of fissile materials\footnote{Fissile materials are
  defined by their ability to sustain a nuclear chain reaction with neutrons of thermal energy, {\it
    e.g.} $^{235}$U and $^{239}$Pu.} is the central concern in nuclear
security, as already recognized in 1946~\cite{acheson}. Under the NPT,
non-nuclear-weapon\footnote{Non-nuclear-weapon states are defined as
  state parties to the NPT that did not manufacture and explode a
  nuclear weapon or other nuclear explosive device before 1 January
  1967.} state parties to the Treaty are required to declare their
``source of special fissionable material in all peaceful nuclear
activities,'' which includes civilian nuclear power production. To
ensure proper accounting of this nuclear material of proliferation
concern, states conclude comprehensive safeguards agreements or
voluntary offer agreements with the International Atomic Energy Agency
(IAEA), where fissile material production is monitored via inspections
and accounting measures. All stages of the nuclear fuel cycle are
subject to IAEA safeguards; this includes: uranium mining, uranium
enrichment, fuel fabrication, use in a reactor, spent nuclear fuel (SNF), and, where
applicable, reprocessing. There are currently 454 operating civilian
nuclear power reactors in the world with dozens more under
construction \cite{WNA}, and thus monitoring of fissile material production
at known nuclear reactor facilities is a key challenge for the IAEA.

An additional challenge in verifying the NPT is confirming that a
nation has declared all of its nuclear material and activities. Such a
task is hindered by the need to continuously verify the absence of
undeclared nuclear reactors, materials, and weapons-relevant
activities. The detection of undeclared nuclear reactors has
historically been largely supported through national technical means,
which involves the collection and analysis of materials, reactor
emanations, and other information by individual states to verify
compliance with international agreements \cite{Stubbs2013}.

Nuclear-related turmoil occurring at the end of the Cold War, including the covert Iraqi nuclear weapons program \cite{Davis1992}, the refusal of North Korea to allow certain IAEA inspections \cite{Hecker2018}, and uncertainty surrounding the status of South Africa's nuclear program \cite{Stumpf1996}, led the IAEA and the international community to recognize that existing safeguards measures failed to provide a complete picture of a state's nuclear
activities. In response, the Model Additional Protocol~\cite{AP2012} was created to supplement existing cooperative IAEA safeguards with
strengthened measures, designed to provide greater assurance for detection of undeclared nuclear materials and activities. The measures
include the incorporation of satellite imagery and other open-source data, and access to information was also increased, through an
expanded scope of reporting, declarations, and complementary access to nuclear sites. The Model Additional Protocol also emphasized a balancing need for non-intrusive monitoring approaches. While the
Model Additional Protocol has already significantly bolstered IAEA safeguards, limitations remain---both procedural and technical---that leave open the possibility that undeclared nuclear reactors go
undetected \cite{Findlay2007}.

The production of nuclear energy results in the generation of
radioactive waste, including spent nuclear fuel (SNF) that has been
removed from the reactor core and any waste materials that remain
after the SNF has been processed. Fission product decays are
present in SNF and reprocessed waste, though at a declining rate
depending upon the amount and age of the material in a storage
facility or repository. The IAEA implements technical verification
measures for the back-end of the nuclear fuel cycle, including SNF storage, reprocessing, and long-term disposition
\cite{pushkarjov}. NPT signatory states are obligated to declare the
uranium and plutonium content of SNF and, currently, thousands of significant quantities (SQs)\footnote{The IAEA
  defines 1 significant quantity (1~SQ) of plutonium as 8 kg of total plutonium provided the \textsuperscript{238}Pu content is less than 80\%.}
of plutonium in SNF are under IAEA safeguards. The majority of
SNF is from light water reactors (LWRs), but the fuel from heavy water-moderated and gas-cooled graphite-moderated reactors also contains plutonium, which may be particularly well suited for nuclear weapons fabrication. The IAEA currently employs containment and surveillance to confirm the presence of the fuel assemblies using,
{\it e.g.} seals on the reactor vessel while the fuel still in use and
seals on dry storage casks when the SNF is sent to permanent
storage. While these approaches may be satisfactory in some scenarios,
they require that the integrity of the items is preserved -- the
so-called continuity of knowledge needs to be maintained.

New international agreements may also shape the safeguards landscape, such as a proposed Fissile Material Cutoff Treaty (FMCT)~\cite{FMCT}. In its most limited version, an FMCT would ban the
production of additional fissile materials---in practice, highly-enriched uranium and separated plutonium---for nuclear weapons. A significant number of countries would support an expanded treaty that would include the reduction of existing stocks of fissile materials available for nuclear weapons by placing agreed-upon quantities of non-safeguarded fissile materials not currently in nuclear weapons under international safeguards. While an FMCT has thus far failed to find political traction, progress towards such an agreement would enhance the need for robust technical means for SNF monitoring and
discovery.

Finally, the Comprehensive Test Ban Treaty (CTBT) bans nuclear
explosions on any scale~\cite{CTBT}. The CTBT was opened for signature
in 1996 and will come into force when 44 specified states that
possessed nuclear reactors as of certain dates in the 1990s have
ratified it. Currently, eight of these states --- China, the DPRK,
Egypt, India, Iran, Israel, Pakistan, and the United States --- have
yet to ratify the treaty. Nonetheless, the CTBT has created a
near-universal global norm against nuclear explosion testing and
international efforts are maintained related to the nuclear explosion
monitoring mission.

\section{Physics of neutrinos from fission sources}
\label{sec:physics}

Nuclear reactors, nuclear explosions, and reactor waste streams
produce neutrinos by the same primary mechanism: nuclear beta
decay. Detection approaches are likewise related, although detection
feasibility varies depending on the source type and distance from
source to detector.

\subsection{Neutrino production in fission sources}
\label{sec:production}

\begin{figure}[t]
    \centering \includegraphics[width=\columnwidth]{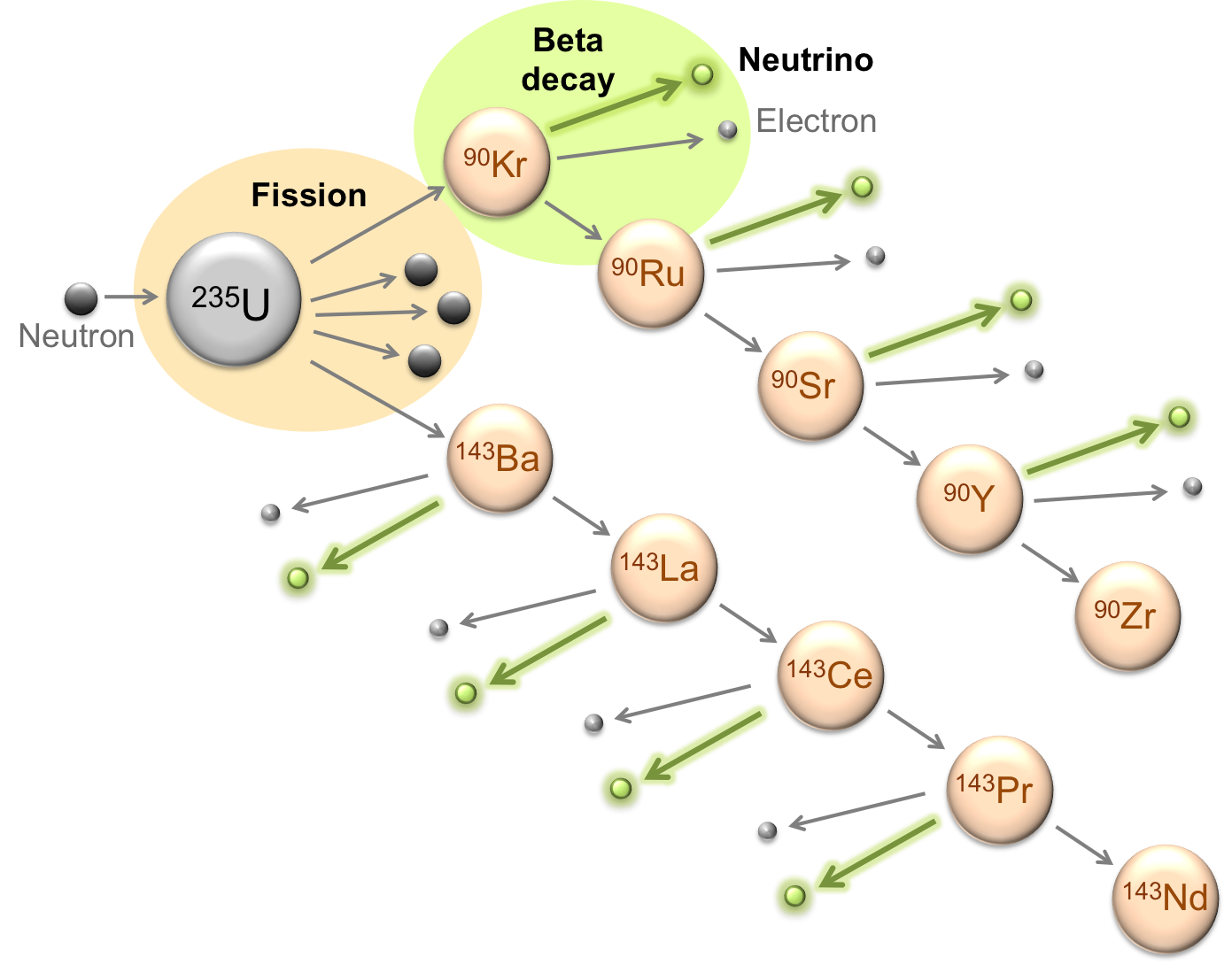}
    \caption{An example of beta decay chain of fission fragments
      resulting in the emission of eight neutrinos.}
    \label{fig:nuprod}
\end{figure}

Neutrinos are produced not by fission itself but the beta decay of
fission fragments.\footnote{Beta decays following neutron capture on materials in a reactor also contribute to the neutrino flux. The effect is small for typical power reactors~\cite{Huber:2015ouo}, but can be significant for certain research reactor configurations~\cite{Ashenfelter:2018jrx}.} Typically, one fission produces two
fragments. Each of these neutron-rich fragments decays an average of
three times. Each decay produces one electron
antineutrino:\footnote{Following common usage, this review uses
  ``neutrino'' as a general term for both neutrinos and
  antineutrinos.}
\begin{equation}
^{A}_{Z}N \rightarrow {^{A}_{Z+1}N'} + e^- + \bar{\nu}_e
\end{equation}
Thus, one fission leads to the emission of roughly six
neutrinos. Figure \ref{fig:nuprod} illustrates this process.

\begin{figure*}[t]
    \centering
    \includegraphics[
      width=0.47\textwidth]{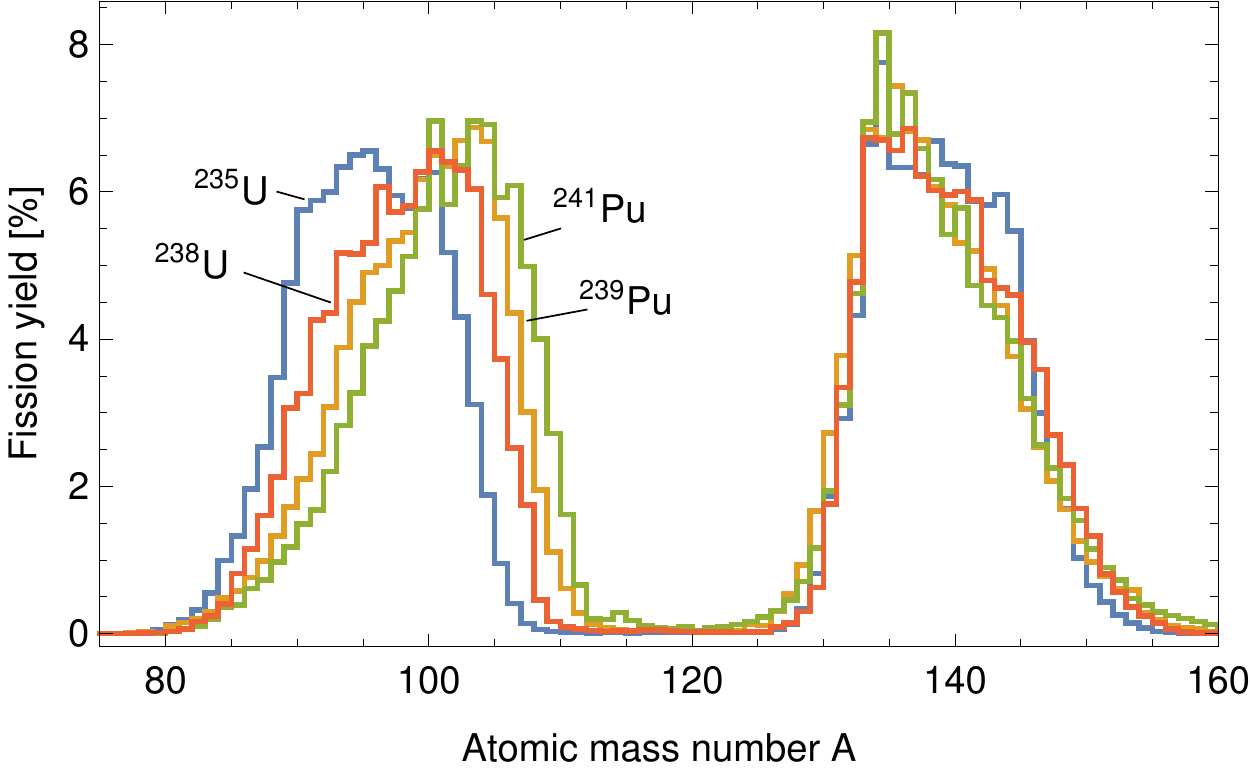}~~\includegraphics[
      width=0.47\textwidth]{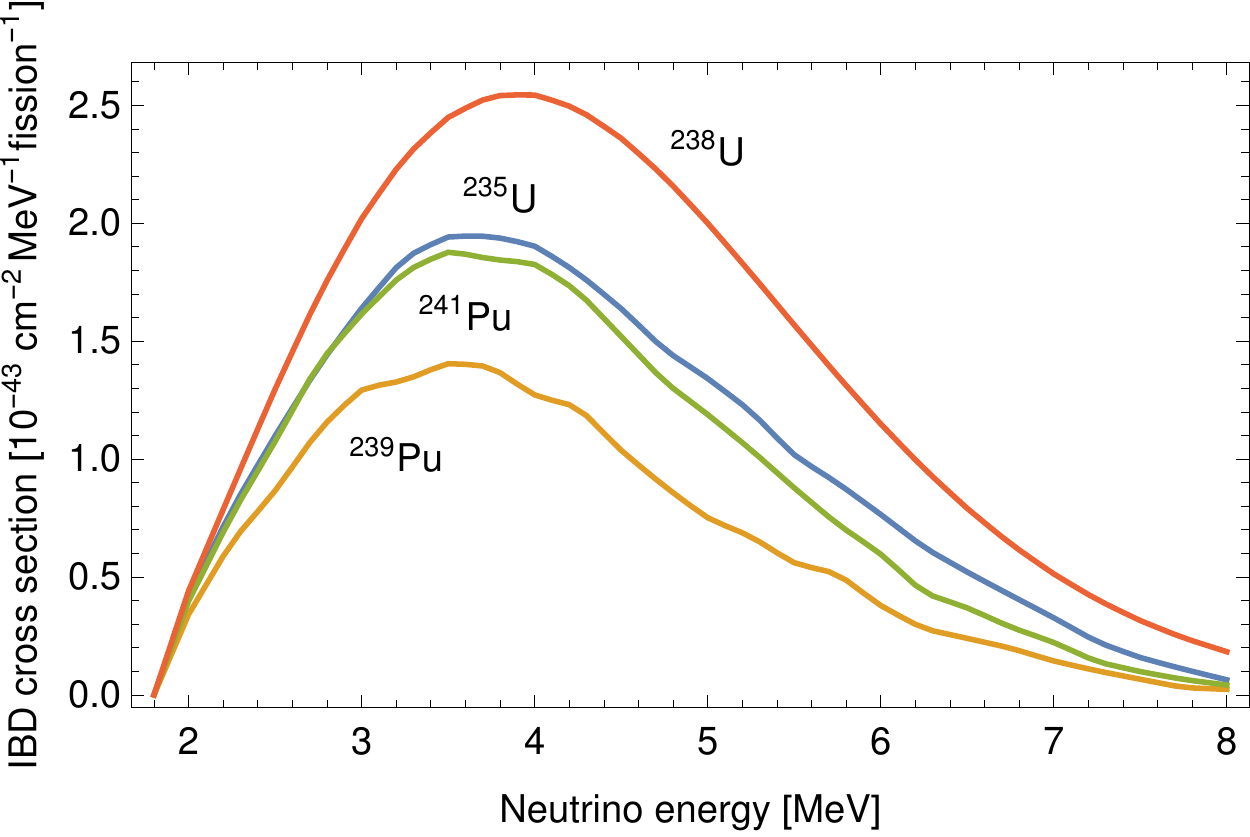}
    \caption{Left: Fission fragment yields from the four major
      nuclides in fission sources according to JEFF~3.3~\cite{jeff}. Right: the detection cross
      section per fission for neutrinos from each of the four fissile
      isotopes, which is obtained as the product of IBD cross section
      and the neutrino flux.}
    \label{fig:fragFlux}
\end{figure*}

Details of the neutrino flux vary according to the nature of the
fission source. Most importantly, the neutrino flux depends on which
nuclides undergo fission, while the energy of the fission-inducing
neutrons has a smaller impact~\cite{Littlejohn:2018hqm}. The dominant
nuclides in most reactors and explosions are $^{235}$U, $^{239}$Pu,
$^{238}$U, and $^{241}$Pu. Neutrino emissions from these nuclides
differ because the fission fragment yields differ. The left side of
Fig. \ref{fig:fragFlux} shows the fission fragment yields. As these
distinct populations of fission fragments decay toward stability, they
give rise to different emission rates and spectra of neutrinos. 
\nb{The right side of Fig. \ref{fig:fragFlux} illustrates how the 
neutrino flux measured via inverse beta decay, a common detection mechanism to be described in Sec.~\ref{sec:detection}, varies between nuclides. Notably, $^{235}$U produces
about 50\% more detectable neutrinos per fission than $^{239}$Pu, with
a harder energy spectrum. }
The neutrino flux from a single source often
includes contributions from fission of multiple nuclides. For example,
in a reactor fueled with low-enriched uranium (LEU), some neutrinos
come from fissions of $^{235}$U and some from fissions of $^{239}$Pu
bred in by neutron capture on $^{238}$U. The overall neutrino flux is
a function of the total fission rate, $R(t)$, the fraction of fissions
occurring on the $k^{th}$ nuclide, $\alpha_k(t)$, and the neutrino
flux from the $k^{th}$ fissioning nuclide, $S_k(E_\nu,t)$, where
$E_\nu$ is neutrino energy and $t$ is time.

Neutrino emissions from a single source often change over time. In a
reactor, the timescale for significant changes in $R$ and $\alpha_k$
(hours to days) is much longer than most of the beta decay lifetimes
(mostly less than a minute). This means that the neutrino flux from a
reactor can be approximated by the equilibrium expression:
\begin{equation}
\label{eq:equilRate}
\phi_\nu^{equil}(E_\nu,t)= R(t) \sum_k \alpha_k(t) S_k(E_\nu,t).
\end{equation}
Eq.~\eqref{eq:equilRate} can also be rewritten in terms of the reactor
thermal power, $P_{th} = R \sum_k \alpha_k E_k$, where $E_k$ is the
mean energy per fission of the $k^{th}$ nuclide:
\begin{equation}
\label{eq:equilPower}
\phi_\nu^{equil}(E_\nu,t)= \frac{P_{th}(t)}{\sum_k \alpha_k(t) E_k} \sum_k \alpha_k(t) S_k(E_\nu,t).
\end{equation}
By contrast, in a nuclear explosion, all fissions occur nearly
instantaneously. The burst-like neutrino emission from an explosion
cannot be approximated by an equilibrium expression. Nonetheless, the
general logic of Eq.~\eqref{eq:equilRate} holds: the neutrino flux
from an explosion is a product of the total number of fissions
(proportional to the fission yield of the explosion) and the sum of
neutrino fluxes from each fission fragment nuclide, weighted by the
fraction of fissions occurring on each nuclide.

Even in a reactor, some notable effects are not covered by the
equilibrium approximation of
Eq.~\eqref{eq:equilRate}--\eqref{eq:equilPower}. One such effect is
the emission of neutrinos from nuclear fuel after the reactor is shut
down or after the fuel is removed. This emission comes from the small
fraction of fission fragments that beta decay over long
timescales. These are the same decays responsible for the long-term
gamma and beta radioactivity of used nuclear fuel. The neutrino rate
from irradiated fuel, whether stored in casks or modified through
chemical reprocessing, is much lower than that from operating
reactors, and the energy spectrum from used fuel is also softer.

\begin{table*}[t]
\caption{Comparison of the three sources of neutrinos discussed in this review.}
\label{tab:sources}
\begin{tabular}{ L{2.5cm} L{2.5cm} L{4cm} L{3cm} L{4.5cm}}
Source & Main origin of $\bar{\nu}_e$ & Time profile of $\bar{\nu}_e$ emission & Energy of $\bar{\nu}_e$ emitted & History of $\bar{\nu}_e$ from this source \\
\hline
\hline
Nuclear reactor & Beta decay of fission fragments & Moderate, quasi-steady state emission over days to months & Up to $\sim 8$ MeV & First detected 1956; millions of interactions detected in many subsequent experiments  \\
\hline
Nuclear explosion & Beta decay of fission fragments & Intense burst over a few seconds & Up to $\sim 8$ MeV, with higher energies emitted earlier & No known detections of $\bar{\nu}_e$ from this source \\
\hline
SNF and fuel reprocessing waste & Beta decay of fission fragments with long lifetimes & Low-level emission that exponentially decays over many years & Up to $\sim 3$ MeV & Likely detected in reactor $\bar{\nu}_e$  experiments but so far indistinguishable from reactor signal and other backgrounds \\ 
\end{tabular}
\end{table*}

Table~\ref{tab:sources} compares the production of neutrinos in the
three sources we consider in this review: reactors, explosions, and
waste streams from reactors.  Recall that the basic production
mechanism is the same for all sources, namely the beta decay of
fission fragments. The energy dependence, time dependence, and
relative intensity of the neutrino flux vary among these three
sources, with implications for the practicality of applications. All
of these sources emit neutrinos isotropically. The fusion reactions
most common in nuclear weapons and the reactions under consideration
for fusion power plants do not produce neutrinos.

\subsection{Basics of detecting fission neutrinos}
\label{sec:detection}

Equations~\eqref{eq:equilRate}-\eqref{eq:equilPower} hint at the
information carried by neutrino emissions from fission sources. To
capture this information, one must observe the neutrinos interacting
in a detector. Consider the generic case of detecting neutrinos some
distance $L$ from a fission source with neutrino flux $\phi_\nu$. Where the spatial extent of the source is small compared to $L$, the
number of detectable neutrino events $N_{det}$ is
\begin{equation}
N_{det}(E_\nu, t) = \frac{\epsilon(E_\nu)}{4\pi L^2} \phi_\nu(E_\nu, t) \sigma(E_\nu) N_{T} P_\textit{surv}(E_\nu, L). 
\label{eq:detect}
\end{equation}
In this expression, $\epsilon$ is the signal detection efficiency,
$\sigma$ is the cross section for the interaction to which the
detector is sensitive, $N_T$ is the number of interaction targets in
the detector, and $P_\textit{surv}$ is the electron antineutrino
survival probability.

Soon after the neutrino was postulated, it was recognized that
neutrino cross sections will be very small and that the most likely
reaction is inverse beta decay (IBD)~\cite{Bethe:1934qn} with a cross
section of  $\approx 10^{-43}\,\mathrm{cm}^2$. The target of this reaction
is a free proton (hydrogen nucleus):
\begin{equation}
\bar{\nu}_e + p \rightarrow e^+ + n.
\label{eq:IBD}
\end{equation}
The threshold for this reaction is $m_n-m_p+2m_{e}\simeq
1.8\,\mathrm{MeV}$ and the visible energy of the positron is given by
$E_\mathrm{vis}=E_\nu - 1.8\,\mathrm{MeV} + 2\times0.511\,\mathrm{MeV}$, that is, there is a
one-to-one correspondence between detected energy and the neutrino
energy $E_\nu$~\cite{Vogel:1999zy}. 
The is correspondenc arises from kinematics:
the energy of the neutrino is carried by the positron and the momentum
by the neutron, where the kinetic energy of the neutron is indeed very
small, on average about $50,\mathrm{keV}$. As a consequence, energy
reconstruction for the neutrino is straightforward but measuring its
direction is difficult. The positron will deposit its energy promptly
and the neutron will thermalize and then capture either on hydrogen or
a specifically added neutron-capture target like gadolinium or lithium;
the neutron-capture elements have a high thermal neutron capture cross
section. This allows to exploit a delayed coincidence between the
prompt positron signal and the delayed neutron-capture signal: both
events happen close in time, $10-200\,\mu\mathrm{s}$, and space,
$5-15\,\mathrm{cm}$. The neutron capture signature can be either emission of gamma rays, in the case of cadmium or gadolinium, or alpha particles and tritons in the case of lithium.  These signatures together form the basis for detector design since
the discovery of neutrinos~\cite{Cowan:1992xc} and greatly suppress
backgrounds from natural radioactivity and cosmic rays. Inverse beta decay on other nuclei besides hydrogen is possible, but generally the cross section is suppressed by nuclear matrix elements and there are fewer targets per unit mass, making hydrogen by far the most practical choice. Suitable
detector mediums contain hydrogen and are transparent: organic
scintillators and water. They both convert the ionization signals of
the positron and neutron capture into light by either scintillation or
Cerenkov radiation.

Interaction modes other than IBD exist: typically they are less practical, but
they can offer certain advantages. In the case of neutrino-electron
scattering,
\begin{equation}
\bar{\nu}~+~e^- \rightarrow \bar{\nu}~+~e^-\,,
\end{equation}
the advantage is that the scattered electron direction may be easier
to reconstruct than the initial momenta of IBD products. This may be
useful for localizing a fission source such as an undeclared
reactor. Backgrounds are often a challenge for this single reaction
product~\cite{Hellfeld:2015xym}. In the case of coherent elastic
neutrino-nucleus scattering (CE$\nu$NS),
\begin{equation}
    \bar{\nu} + N \rightarrow \bar{\nu} + N\,,
\end{equation}
 one advantage is that the cross section is coherently enhanced by the
 contribution of all neutrons in the target
 nucleus~\cite{Freedman:1973yd}. For a large nucleus such as germanium
 or xenon, the enhancement is two orders of magnitude over IBD per
 unit detector mass. Another advantage is that CE$\nu$NS has no
 kinematic threshold, so neutrinos below the IBD threshold of 1.8\,MeV
 are in principle observable.  For CE$\nu$NS, the primary difficulties
 are detecting the very low-energy nuclear recoil, typically
 $\mathcal{O}(10-100)$\,eV, and suppressing background in this
 low-energy range. Owing to these small recoil energies, this reaction
 has been observed only recently~\cite{Akimov:2017ade}, albeit using
 neutrinos from a pulsed source with about 10 times higher average
 energy than reactor neutrinos.

The final component of Eq.~\eqref{eq:detect} accounts for neutrino
flavor oscillation. This is the quantum mechanical phenomenon that
allows a neutrino created in one flavor (electron, muon, or tau) to be
detected as a different flavor~\cite{NobelArt,KajitaNobel}. Fission
sources produce only electron antineutrinos, and IBD is sensitive only
to this flavor. When electron antineutrinos propagate, some of them
become invisible to IBD detectors as they oscillate into non-electron
flavors; only the surviving electron antineutrinos are observable. One
upside of oscillations is that $P_\textit{surv}$ has a nonlinear
dependence on $L$, the distance from source to detector. Thus
oscillations can break certain degeneracies~\cite{Jocher:2013gta}. The
more essential upside is that neutrino oscillations are a major focus
of basic research. The presence of oscillations in
Eq.~\eqref{eq:detect} has made reactors a key source for fundamental
physics experiments. These experiments have played a critical role in
developing technology that may be used for neutrino applications.

Neutrinos interact only via the weak force, and thus, neutrino cross
sections are very small in absolute terms. Consequently, neutrino
detection requires careful control and reduction of potential
background sources. Common strategies are: selection of radio-clean
construction materials; use of engineered shielding against neutrons
and gamma rays; locating the experiment underground; particle
identification; spatial segmentation. For a more detailed discussion,
which is beyond our scope, see for instance~\cite{Bowden:2012um}.


\subsection{Information content of fission neutrino signals}
\label{sec:info}

The information contained in fission neutrino signals is described by
Eqs.~\eqref{eq:equilRate}--\eqref{eq:detect}: substituting
Eq.~\eqref{eq:equilPower} into Eq.~\eqref{eq:detect}, and suppressing
the energy and time dependence for simplicity, yields
\begin{align}
N_{det} = \left( \frac{\epsilon N_T \sigma}{4\pi} \right) 
\left( \frac{P_{e \rightarrow X}(L)}{L^2} \right)
\frac{P_{th}}{\sum_k \alpha_k E_k} \sum_k \alpha_k S_k.
\label{eq:info}
\end{align}
The first factor in parentheses contains parameters which the detector
operator can determine. The last parameter, $S_k$, is also fairly well
known for the major nuclides, when fissioned by thermal
neutrons.

In this context it is necessary to point out that reactor antineutrino
fluxes have been subject of intense scrutiny since 2011, when two new
evaluations were
conducted~\cite{Mueller:2011nm,Huber:2011wv} that up-corrected the
resulting IBD rates by approximately 6\%. This in turn gave rise to
the to the so-called reactor antineutrino anomaly
(RAA)~\cite{Mention:2011rk}: \nb{all past measurements, which had been interpreted as being in agreement with prior flux predictions, now indicated a significant rate deficit relative to those more modern updates.} One possible solution could be the existence
of a fourth, so-called sterile neutrino, which triggered considerable
experimental activity~\cite{Abazajian:2012ys} and to date remains a
viable possibility~\cite{Dentler:2018sju}. The RAA and other
discrepancies in prediction and measurements of the neutrino spectrum
are under active study, for a review see~\cite{Hayes:2016qnu}, and it
is clear that for applications these issues need to be resolved by
experimental measurement. Therefore, calibrating reactor antineutrino
fluxes from a range of different reactors at different stages in their
fuel cycle is a mandatory, and entirely feasible, ingredient for this
application. As an example consider the recent measurement of the neutrino yield spectrum from uranium-235 and plutonium-239 by the Daya Bay collaboration~\cite{Adey:2019ywk}.

The other factors depend on information which the detector operator
may not know: the distance $L$ to the reactor (unknown if, for
example, the reactor is hidden), the reactor power level $P_{th}$, and
the fission fractions $\alpha_k$ inside the reactor core. Evidently,
by observing neutrino emissions from the reactor, one can possibly
infer a combination of:
\begin{itemize}
\item How far away the reactor is;
\item What power level the reactor is operating at; and
\item What the reactor is burning for fuel.
\end{itemize}
These pieces of information can be in principle distinguished using
the time and energy dependence of the observed neutrino
flux. Furthermore, one or more of the above source characteristics may
be constrained by non-neutrino data or by declared reactor operating
histories. In this case, a combined analysis of neutrino and
non-neutrino data could further disentangle the components above. The assumption is that for deployments under cooperative safeguards the distance to the reactor is known at the percent-level.

In a similar manner, the neutrino signal from a nuclear explosion
carries information about how far away the explosion occurred, how
much fission yield the explosion contained, and which nuclide was used
as a nuclear explosive. Neutrino emissions from SNF carry some
information about the fuel location and time elapsed since the fuel
has been discharged from the reactor. However, as we describe in
Secs.~\ref{sec:near}-\ref{sec:weapons}, collecting this information is
more practical near reactors than from waste streams or explosions. To
give context for those comparisons, Sec.~\ref{sec:history} describes
the history of neutrino detection at fission sources.

\section{History of fission neutrino detection}
\label{sec:history}

\subsection{Fundamental physics: first detection and neutrino oscillation experiments}

The first detection of a neutrino of any kind occurred at a nuclear
reactor. In the 1950s, a team led by Frederick Reines and Clyde Cowan
observed neutrino emission from a plutonium production reactor at the
US Atomic Energy Agency (now Department of Energy) Savannah River
site~\cite{Cowan:1992xc}. The Cowan-Reines detector was small
($<$0.5~ton), but its use of organic scintillator, doping, and
segmentation established design principles that remain in use 60 years
later. Over five million neutrinos have now been detected at nuclear
reactors around the world. Physicists, including Reines, considered
making basic physics measurements using nuclear weapon tests as a
source~\cite{ReinesNobel}. To date, however, no neutrinos from nuclear
explosions have been observed. Neutrinos from SNF make some
contribution to data sets collected at nuclear power plants, but that
component is not statistically distinguishable from the much larger
contribution from operating reactors. Reactors remain the only fission
source from which neutrinos have been conclusively detected.

As the brightest neutrino sources on Earth, nuclear reactors have
attracted particle physicists over many decades for dozens of
fundamental studies. Early experiments used ton-scale detectors
located within a few tens of meters of reactor cores. Efforts
searching for evidence of neutrino oscillation were mounted in the
1970s through the 1990s in the
USA~\cite{Reines:1980pc,Greenwood:1996pb,Riley:1998ca},
France~\cite{Kwon:1981ua,Cavaignac:1984sp,Declais:1994su},
Switzerland~\cite{Zacek:1986cu} and the
USSR~\cite{Kuvshinnikov1991,Vidyakin1994}.  In the late 1990s, the
Chooz~\cite{Apollonio:1999ae} and Palo Verde~\cite{Boehm:2001ik}
experiments extended the baseline for reactor neutrino oscillation
searches to $\approx1$\,km using detectors of 10\,ton scale.  In the
early 2000s, the KamLAND experiment in Japan used a kiloton-scale
liquid scintillator (LS) detector to observe neutrinos from nuclear
reactors over 100\,km away~\cite{Eguchi:2002dm}. The energy-dependent
deficit of electron antineutrinos seen by KamLAND, a consequence of
flavor oscillations, helped to establish that neutrinos have
mass. More recently, LS detectors on the 10\,ton scale have made
precision oscillation measurements at distances in the range of
$400-1900$\,m from nuclear power plants in China~\cite{An:2012eh},
Korea~\cite{Ahn:2012nd}, and France~\cite{Abe:2011fz}. Beyond measuring fundamental neutrino parameters, these recent experiments provided stringent tests of the reactor neutrino emission models by performing high precision energy spectrum measurements. 

The fundamental physics experiments described above laid the
foundation for possible reactor monitoring applications using neutrino
emissions. They provide detection capability demonstrations at
standoff distances spanning the near field and far field, while also
developing an understanding of reactors as a neutrino source and the
important background mechanisms that limit sensitivity.

\subsection{Application-oriented experiments}

\begin{figure}[t]
\includegraphics[width=0.9\columnwidth]{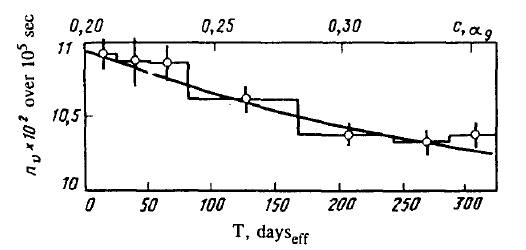}
\includegraphics[width=0.9\columnwidth]{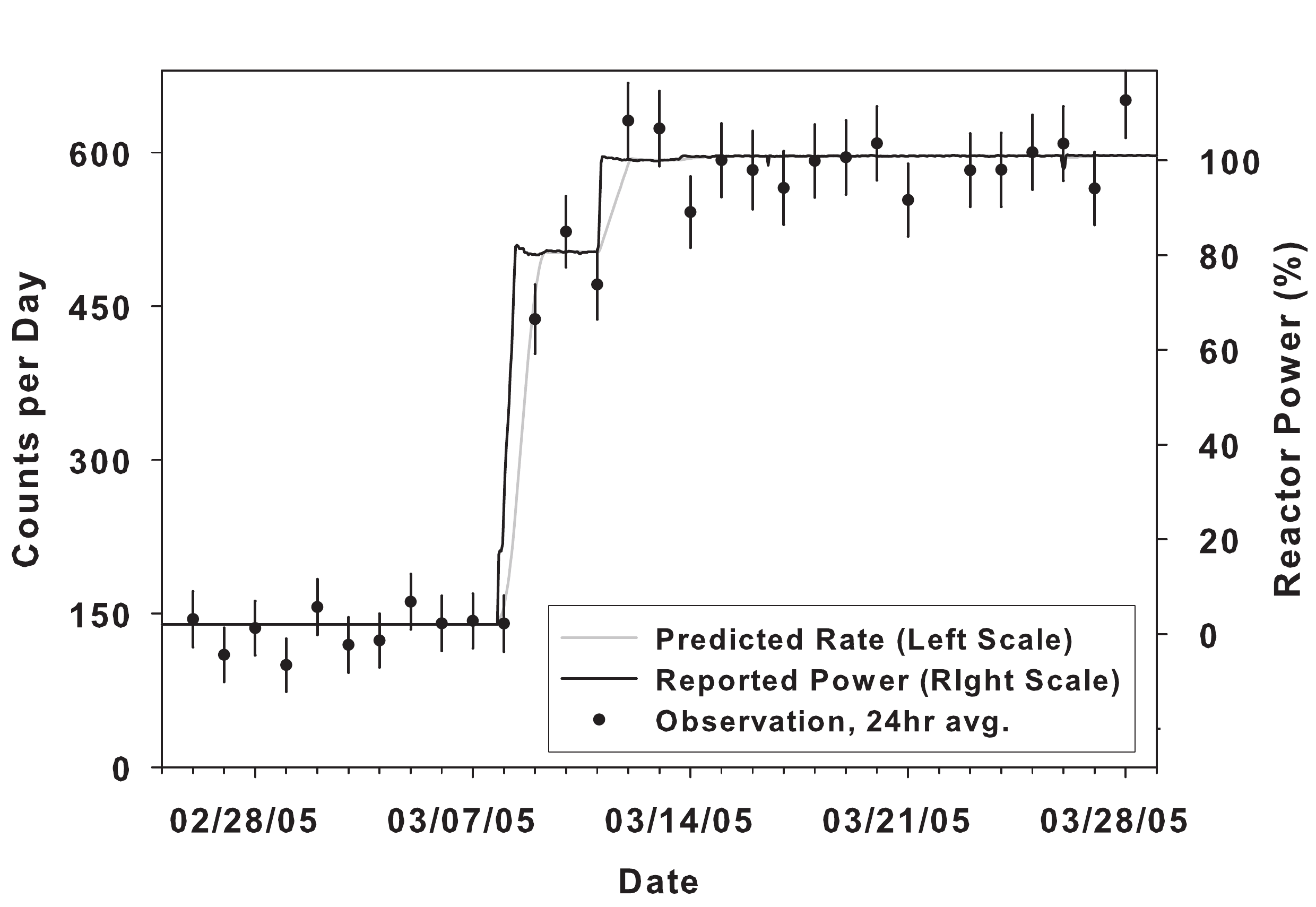}
\caption{\label{fig:burnup-demo} (Top) Measurement of fuel burnup at
  Rovno. The detected reactor antineutrino rate decreases over an
  observation period of 300\,days as production and burning of
  $^{239}$Pu reduces the emitted antineutrino flux, figure
  from~\cite{Klimov:1994}. (Bottom) Observation of reactor start-up at
  SONGS. The correlated event rate tracks the change in reactor power
  at start-up, where the events measured at zero power are due to
  background, figure from~\cite{Bowden:2008ih}.}
\end{figure}

That reactor neutrinos could be useful for nuclear security, was
recognized in 1978 by L. Mikaelyan and
A. Borovoi~\cite{Borovoi:1978,Mikaelian}. Several demonstrations of
the reactor monitoring concept have been performed in the very
near-field range, $7-25$\,m from reactors. Pioneering work was
undertaken in the 1980s at the Rovno power plant in the former Soviet
Union~\cite{Klimov:1994}. This demonstration used a 0.5-ton, Gd-doped
LS (GdLS) detector deployed in a below-ground gallery about 20~m from
the reactor core. This high-efficiency detector recorded almost 1000
neutrino interactions per day with a signal-to-background (S:B) ratio
considerably greater than unity. Over several years, this group
demonstrated rapid determination of reactor on/off state transitions,
tracking of reactor power levels, and measurements of the change in
neutrino rate and spectrum due to fuel evolution (burnup), see
Fig.~\ref{fig:burnup-demo}.

The next effort to focus on reactor monitoring was based at the San
Onofre Nuclear Generating Station (SONGS) in the United
States. Beginning in the early 2000s, physicists from the Lawrence
Livermore and Sandia National Laboratories constructed and deployed
several neutrino detectors. The goal was to demonstrate that simple
designs could operate unattended for long periods, collecting neutrino
data suitable for reactor monitoring. The 0.6-ton, GdLS SONGS1
detector was deployed in a below-ground gallery about 20~m from the
reactor~\cite{Bowden:2006hu}. The device was calibrated automatically
and maintained stable operation from 2003 till 2008. The simple design
yielded a modest efficiency, with about 500 IBD events recorded per
day. Analysis of the SONGS1 data set produced monitoring demonstrations
similar to those achieved at Rovno: reactor
state~\cite{Bowden:2008ih,Bernstein:2008tj}, reactor power
~\cite{Bernstein:2008tj}, and the rate change due to fuel
burnup~\cite{Bowden:2008gu} (Fig.~\ref{fig:burnup-demo}). This group
also developed a more optimized homogeneous GdLS detector
design~\cite{Classen:2014uaa} with improved detection efficiency and
energy resolution.

The Nucifer collaboration~\cite{Boireau:2015dda}, based in France,
performed a monitoring demonstration at the 70\,MW$_\mathrm{th}$ OSIRIS research
reactor. The aim was to develop a detection system suitable for
operation within a research reactor building. Considerable effort went
into the certification process that allowed the detector to operate
within 7\,m of a reactor core. The design was based on 0.8\,tons of
GdLS in a single vessel. Significant shielding was required to
suppress reactor-correlated $\gamma$-ray backgrounds. At the
relatively modest overburden of 12~mwe\footnote{mwe is short for meter
  water equivalent and allows to express overburden independent of the
  specific rock/soil composition.}, the use of PSD\footnote{PSD stands
  for pulse shape discrimination, which allows to distinguish
  particles based on their mean energy loss per traveled distance,
  $\mathrm{d}E/\mathrm{d}x$. Particles with a high
  $\mathrm{d}E/\mathrm{d}x$ tend to produce a broader light emission
  pulse than particles with small $\mathrm{d}E/\mathrm{d}x$, like beta
  rays~\cite{BROOKS1959151}.}%
capable GdLS was important for
suppression of cosmogenic correlated neutron backgrounds. Recording
almost 300\,IBD interactions per day with S:B = 1:4, Nucifer was able
to follow the operation state and power level of the OSIRIS reactor.

Subsequent efforts addressed the desire to operate detectors on the
earth's surface without cosmic-ray attenuating overburden, since this
enables deployment in a much broader range of locations. Particle type
identification and interaction localization capabilities are key
design features that have been developed to address the much greater
background encountered at the earth's surface. 
Examples of such techniques include segmentation, which provides position resolution roughly equivalent to the segment pitch in compact detectors, and neutron capture identification based on event topology and/or incorporation of $^6$Li, which yields a tightly localized signal upon neutron capture.
The event localization capability provided by segmentation allows selections based on spatial correlations, in addition to the timing correlation supplied by the IBD reaction. For example, use of  event location information to require a spatial coincidence between the prompt and delayed components of an IBD event candidate is effective at suppressing random temporal coincidences of singles backgrounds. The spatial pattern (topology) of energy depositions within the prompt and delayed components themselves can also be of use. Examples include attempts to preferentially select deposition patterns corresponding to  IBD positrons (primary positron ionization energy loss and the Compton scattering of the resulting $511$\,keV annihilation $\gamma$-rays) and neutron captures on Gd (Compton scattering of multiple MeV-scale $\gamma$-rays).

The PANDA project~\cite{Kuroda:2012dw,Oguri:2014gta} realized several
generations of detectors based on an heterogeneous arrangement of
plastic scintillator (PS) and Gd coated sheets. Operation of the
PANDA-360 prototype at a reactor in Japan without overburden provided
a low significance hint of reactor state determination with S:B of
less than 1:15~\cite{Oguri:2014gta}. Similar approaches have been
pursued by groups in India~\cite{Mulmule:2018efw} and the United
Kingdom~\cite{Carroll:2018kad}.

The group responsible for SONGS1 developed an approach that provides a
distinct neutron capture identification signal using $^6$LiZnS neutron
capture screens~\cite{KIFF2011412}. When layered between segmented PS
bars, highly localized neutron captures on $^6$Li could be identified
via the slow ZnS scintillation time constant using PSD. This approach
strongly suppresses background events due to spallation processes that produce multiple neutrons that can enter a detector and be captured with a time correlation structure similar to IBD~\cite{Bowden:2012um} that are difficult to identify in
detectors that use Gd or other $\gamma$-ray emitting neutron capture
agents, while also reducing accidental coincidence backgrounds. A
small prototype deployed in a 20\,ft ISO shipping container at SONGS
without overburden did not have sufficient sensitivity to observe
neutrinos, but did demonstrate powerful background
reduction~\cite{Reyna-inmm-segmented}. The use of wavelength shifting
(WLS) materials to efficiently transport $^6$LiZnS scintillation to
photo-sensors at the edges of a heterogeneous detector arrangement,
first developed for neutron scattering
experiments~\cite{VANEIJK2004260}, is an important element of this
approach.

As discussed in Sec.~\ref{sec:far}, demonstrations of far-field
capabilities beyond ten kilometers or so require kiloton-scale
detectors, with target masses increasing to the megaton scale beyond
$\sim\!100\!-\!200$ kilometers. The first dedicated far-field
demonstration of reactor monitoring has been initiated by the US-UK
WATCHMAN collaboration \cite{Askins:2015bmb}. WATCHMAN is an acronym
for the WATer CHerenkov Monitor of ANtineutrinos, a Gd-doped water
Cerenkov detector with a fiducial mass of 1000\,tons, located in an
underground site 25\,km from a dual-reactor complex in the UK. The WATCHMAN collaboration currently plans for start of data-taking operations in approximately 2025.

\subsection{Return to fundamental physics with near-field reactor observations}

In recent years, searches for new physics in the neutrino sector have
brought basic science attention back to near-field reactor
observations. In 2011, recalculations of reactor neutrino fluxes were
found to be significantly higher than the ensemble of
observations~\cite{Mueller:2011nm,Mention:2011rk,Huber:2011wv}. Among
other possibilities, this discrepancy could be explained by the
existence of a sterile neutrino, a neutral fermion with even weaker
couplings to matter than the Standard Model neutrinos or by deficiencies in the nuclear data and methods used to  predict the reactor antineutrino flux. Indeed, the discrepancy between recent precision energy spectrum measurements~\cite{An:2015nua,RENO:2015ksa} and prediction, most prominent near 5\,MeV, is a strong indication that such deficiencies exist.  

A wide variety of detector designs have been proposed to test the
sterile neutrino hypothesis. Many of these detectors must operate at
or near the surface with limited cosmic ray attenuating overburden due
to the configuration of the host reactor facilities, and are designed
to provide good energy resolution, detection efficiency, and/or
background rejection.  Here we detail some effort of particular
relevance to reactor monitoring applications.

The NEOS experiment~\cite{Ko:2016owz} operates in a below-ground
location similar to SONGS1. Using a $1$\,ton GdLS target and PSD for
background suppression, a signal-to-background of 20 and event rate of
$\sim$2000 IBD interactions per day are achieved. In the context of
near-field reactor monitoring, this device provides high statistics
for rapid determination of reactor status, power level, and
measurement of the reactor antineutrino energy spectrum. NEOS
represents an excellent example of what can be achieved using a modern
GdLS material in a location with $20$\,mwe or more overburden. The STEREO~\cite{Almazan:2018wln} and Neutrino-4~\cite{Serebrov:2018vdw} experiments have also successfully performed reactor antineutrino measurements at research reactors using GdLS target material. In both cases, modest overburden of order $10$\,mwe was available. DANSS~\cite{Alekseev:2018efk} has also achieved a high reactor antineutrino counting rate using a heterogeneous detector composed of PS bars and Gd coated sheets. Operating in a location beneath a power reactor core, DANSS enjoys high antineutrino flux and $\sim 50$\,mwe overburden, providing sufficient sensitivity to observe small flux variations due to reactor operations~\cite{Alekseev:2019yay}.

\begin{figure}[t]
\centering
\includegraphics[width=\columnwidth]{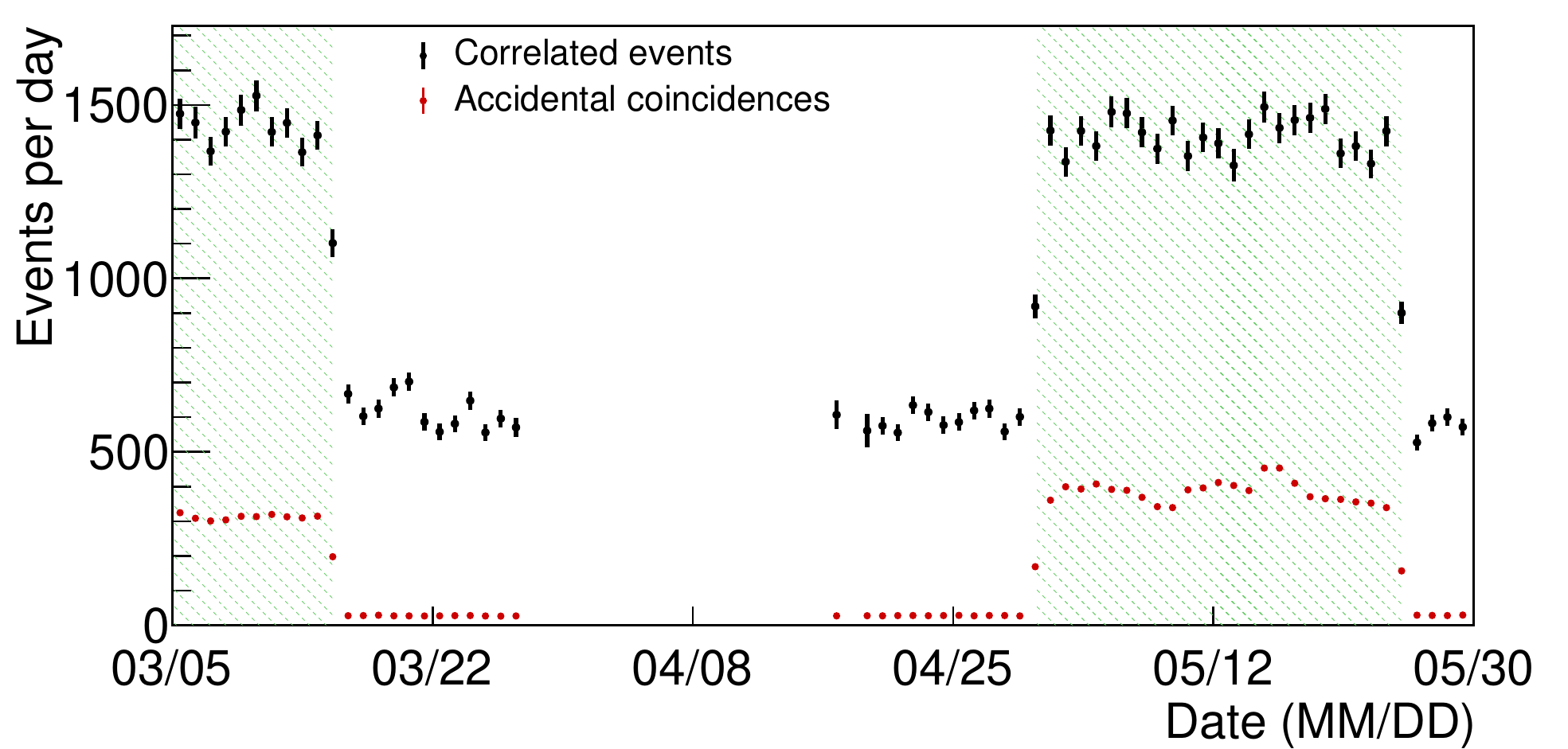}
\caption{\label{fig:surface-demo} On-surface measurement of reactor
  operational state by PROSPECT with less than 1~mwe of
  overburden. Green (gray) shaded periods correspond to full power operation
  of the host reactor, with the correlated event excess relative to
  reactor off periods being due to detection of reactor
  antineutrinos, figure from~\cite{Ashenfelter:2018iov}. }
\end{figure}

The PROSPECT experiment ~\cite{Ashenfelter:2015uxt} has made a
significant advance by performing the first demonstration of
on-surface reactor antineutrino detection with S:B $\sim 1$
(Fig.~\ref{fig:surface-demo}), this being achieved at a research
reactor facility with less than 1\,mwe
overburden~\cite{Ashenfelter:2018iov}.  This result can now serve as a
benchmark for reactor monitoring use cases involving on-surface
detector deployment, \textit{e.g.}~\cite{Carr:2018tak}.  The PROSPECT
detector design incorporates multiple capabilities that combine to
efficiently reject cosmogenic backgrounds.  The use of 4\,tons of PSD
capable $^6$Li-doped LS (LiLS) provides fast neutron and neutron
capture identification, while a 2D segmented geometry ($14.5$\,cm
pitch) provides event localization and topology. An emphasis on
efficient, uniform light collection results in very good energy
resolution for an organic scintillator
detector~\cite{Ashenfelter:2018cli}, which has been utilized in a
measurement of the $^{235}$U reactor antineutrino energy
spectrum~\cite{Ashenfelter:2018jrx}.  Initial background predictions
for PROSPECT~\cite{Ashenfelter:2015uxt} are in good agreement with the
data reported in~\cite{Ashenfelter:2018iov}, including observation of
spectral features due to multiple neutron and neutron inelastic
processes.

Several other approaches focus on more finely-grained segmentation
than PROSPECT. SoLid was among the first near-field reactor efforts to
propose and realize finer-grained three-dimensional segmentation as a
background rejection strategy~\cite{Abreu:2018ajc,Abreu:2018pxg}.
This detection concept combines $^6$LiZnS neutron capture sheets,
$5$\,cm cubes of PS, and WLS optical fibers, providing 3D topological
information and neutron capture identification.  SoLid has collected
reactor data and analysis is ongoing to determine the extent to which
event topology information obtained from relatively fine grained 3-D
segmentation can be used to reject fast neutron backgrounds in a
ton-scale detector.  The goal is to identify positron-like event
topologies including spatially isolated depositions from $511$\,keV
annihilation gamma rays.

NuLat uses a light collection arrangement known as the Raghavan
Optical Lattice (ROL) to obtain fine-grained 3-D segmentation (also
$\sim5$\,cm pitch) and efficient light collection~\cite{Lane:2015alq}.
The use of homogeneous $^6$Li-doped materials in combination with the
ROL promises access to all proposed particle ID methods simultaneously
-- fast neutron recoil PSD, neutron capture PSD, and fine-grained
topological information -- and therefore should have excellent
background rejection. The current availability and optical performance
of $^6$Li doped PSD-capable plastic scintillators has limited the
extent to which the concept has been demonstrated to date.

Inspired by the SoLid, SNL/LLNL, and NuLat segmented efforts, CHANDLER
uses $^6$LiZnS screens, wavelength shifting plastic scintillator, and
an ROL to provide fine grained topology information, a distinct
neutron capture tag and good optical collection and energy resolution
compared to SoLid. In contrast to NuLat, the CHANDLER concept can be
realized with materials that are readily available from commercial
vendors. As with SoLid, the ability to identify and reject background
is based on event topology information obtained from relatively fine
grained segmentation in combination with a distinct neutron capture
tag. CHANDLER reports IBD detection including the spectrum from
several months operation without overburden at a 2900 MW$_{th}$ pressurized water reactor
using an 80\,kg miniCHANDLER
prototype~\cite{Haghighat:2018mve}. CHANDLER is among the efforts to
have demonstrated a main advantage of solid plastic detectors: the
miniCHANLDER prototype is mounted inside a road-legal trailer, it can
be driven to the deployment site and data taking can start within
hours of deployment.

\section{Applications to known reactors: fissile material production monitoring}
\label{sec:near}

\subsection{Existing approaches}
\label{sec:reactor_existing}

The IAEA implements a variety of technical measures to verify a state
is in compliance with its safeguards agreements. Safeguards are
primarily designed to detect the diversion of nuclear material from
declared facilities, undeclared processing or production of nuclear
materials at declared facilities, and undeclared facilities processing or producing nuclear material. The IAEA implements safeguards using a
combination of nuclear material accountancy, nondestructive and
destructive measurements, and containment and surveillance.

Measurements of nuclear material confirm the declared mass and
composition of the material, typically by employing nondestructive
measurements, {\it e.g.}, measuring the weight of a uranium sample
using a scale, and measuring its isotopic composition using gamma
spectroscopy. Destructive measurements are employed when necessary,
{\it e.g.}, measuring the isotopic composition of a solution of
dissolved irradiated fuel using mass spectrometry. Measurements also
verify the declared operation of a process, {\it e.g.}, by measuring
the flow rate of UF\textsubscript{6} in a gas centrifuge
plant. Furthermore, environmental sampling \ph{at pre-designated
  locations within declared facilities} is frequently applied to
detect the presence of undeclared materials or declared materials in
anomalous locations, which can be indicative of diversion, undeclared
processing. \ph{Wide-area sampling, {\it i.e.} outside of declared
  facilities, is permitted under the Additional Protocol to uncover
  undeclared facilities, however it is not approved as a routine
  inspection tool and usually reserved for cases where a specific
  concern exists.}

Finally, containment and surveillance \ph{are the key technologies} to
detect undeclared access to and/or movement of nuclear
material. Containment is implemented using tamper-indicating seals
applied to nuclear material containers and process controls; attempts
to access or move the nuclear material, or change the operation of a
process, would be detected if the integrity of seals were
compromised. Surveillance is primarily implemented using cameras to
observe material balance areas and process controls. Currently, most
safeguards surveillance systems do not provide real-time remote
monitoring; however, the IAEA is working to transition its
surveillance systems to provide real-time remote monitoring of many
facilities in the near future.

The \ph{declared burnup of spent fuel} is primarily verified
using accountancy of the fresh and irradiated fuel and
nondestructive analysis of the fresh and irradiated
fuel. Nondestructive analysis of the fresh fuel serves to verify its
declared mass and enrichment, which is accomplished by relatively
simple weight and gamma spectroscopic measurements.  However,
nondestructive analysis of the irradiated fuel does not yield a direct
measurement of the fuel's isotopic composition, including the fuel's
residual uranium content and the plutonium bred in the fuel during
irradiation, because gamma and neutron emissions by fission products
in the fuel mask radiation emissions from the uranium and plutonium
isotopes.

Radiation measurements of SNF are used to confirm that it is
consistent with the declared \ph{initial enrichment, burnup and
  cooling time. The most widely used technique, is based on measuring
  the Cerenkov radiation emanating from SNF within the water of the spent
  fuel pond. This is accomplished using the so-called Cerenkov Viewing
  Device (CVD)~\cite{CVD}, which essentially just confirms that the
  SNF is present and exceeds a certain level of overall
  radioactivity. The advantages of the CVD are that it is fast, it does
  not require fuel movement and does not get into contact with the pool
  water.} The fission and/or activation product content of the fuel
\ph{can be} measured using gamma spectroscopy and/or neutron
coincidence counting\ph{, but these techniques are rarely
  employed}~\cite{IAEA2011}. Except in the case of a few research
reactors, typically with a thermal power in excess of 25\,MW,
safeguards do not implement real-time monitoring of reactor
operations. For those exceptional reactors power is measured by using
the advanced thermohydraulic power monitor~\cite{atpm}, where the flow
rate of coolant and temperature rise across the reactor are measured.
\ph{For SNF in dry storage the default technologies are tamper indicating
seals and surveillance.}  The majority of nuclear reactors has a significant amount of fertile material, {\it i.e.} material that under neutron-irradiation can become fissile, present in the reactor core; in power reactors uranium-238 is the most important of those. As a consequence, these reactors produce some fissile material, notably plutonium-239, during operation. The amount and quality of plutonium produced is a function of the total burn-up and the initial fuel enrichment and composition: for a typical 3\,GW$_\mathrm{th}$ pressurized water reactor a plutonium production rate of 100--200\,kg per year is not unusual. Therefore, verifying burn-up, enrichment and fuel composition is an important part of safeguards. In particular, a willful mis-declaration of any of those quantities would allow for the production of excess plutonium (or a more weapons-usable grade) or to overstate the amount of plutonium which is consumed. The latter is critical for international agreements to reduce the stockpile of fissile material.

Real-time remote monitoring of nuclear reactor operations has been
demonstrated using satellite and aerial imagery of heat signatures
emanating from the reactor. The reactor's thermal output, either in
terms of its injection of hot water into a reservoir, or its emission
of warm air from its cooling towers, can reveal the on/off state of
the reactor, and can be correlated to the reactor's operating
power~\cite{Garrett2010,Lee2015}. At shorter ranges ({\it e.g.}, hundreds of
meters), sky shine (gammas scattering in the air above a reactor
containment building) can also reveal the on/off state of the reactor
\cite{Wahl2016}.

\subsection{Neutrino-based approaches}

Section~\ref{sec:info} contains a description of how neutrino emissions
carry information about reactor power levels and fuel contents. This
information, collected in real time, could complement existing reactor
monitoring techniques. The basic neutrino observables are neutrino
rate, neutrino energy spectrum, and time evolution of neutrino
spectrum and rate. These observables in turn allow, at least in
principle, to measure the fission rates, $f_I(t)$, and thus, also
reactor power. The rate at which the fission rate, $f_I(t)$, changes
with time is indicative of the initial fuel enrichment. All neutrino
observations are measuring the neutrino emission from the entire
reactor core and thus any inferred quantity always represents a core
average. \ph{That is, neutrino-based technology provides a form of
  bulk accountancy, whereas current procedures are mostly providing
  item accountancy. In the context of some advanced reactor designs,
  like molten salt reactors, item accountancy will not be possible,
  providing additional motivation for neutrino-based approaches.}

Reactors with a high neutron flux density will produce more fissions
per unit mass of the fissile nuclide. This relationship connects
neutrino measurements to the core fissile inventory. Smaller reactors
contain less plutonium and thus it is easier to achieve an absolute goal like detection of 1\,SQ.
This indicates that commercial, multi-GW light water moderated reactors are a challenging target for neutrino safeguards relative to the IAEA goals.
However, even for those reactors, neutrino safeguards can provide a 1--2\% core-wide plutonium inventory, which exceeds the accuracy of any other practical approach; a capability which would become relevant in the context of the FMCT. On the other hand, for typical plutonium
production reactors, research reactors, and small modular reactors
neutrinos can meet the IAEA goals both in terms of quantity and
timeliness of the result.

\begin{figure}[t]
    \centering \includegraphics[width=\columnwidth]{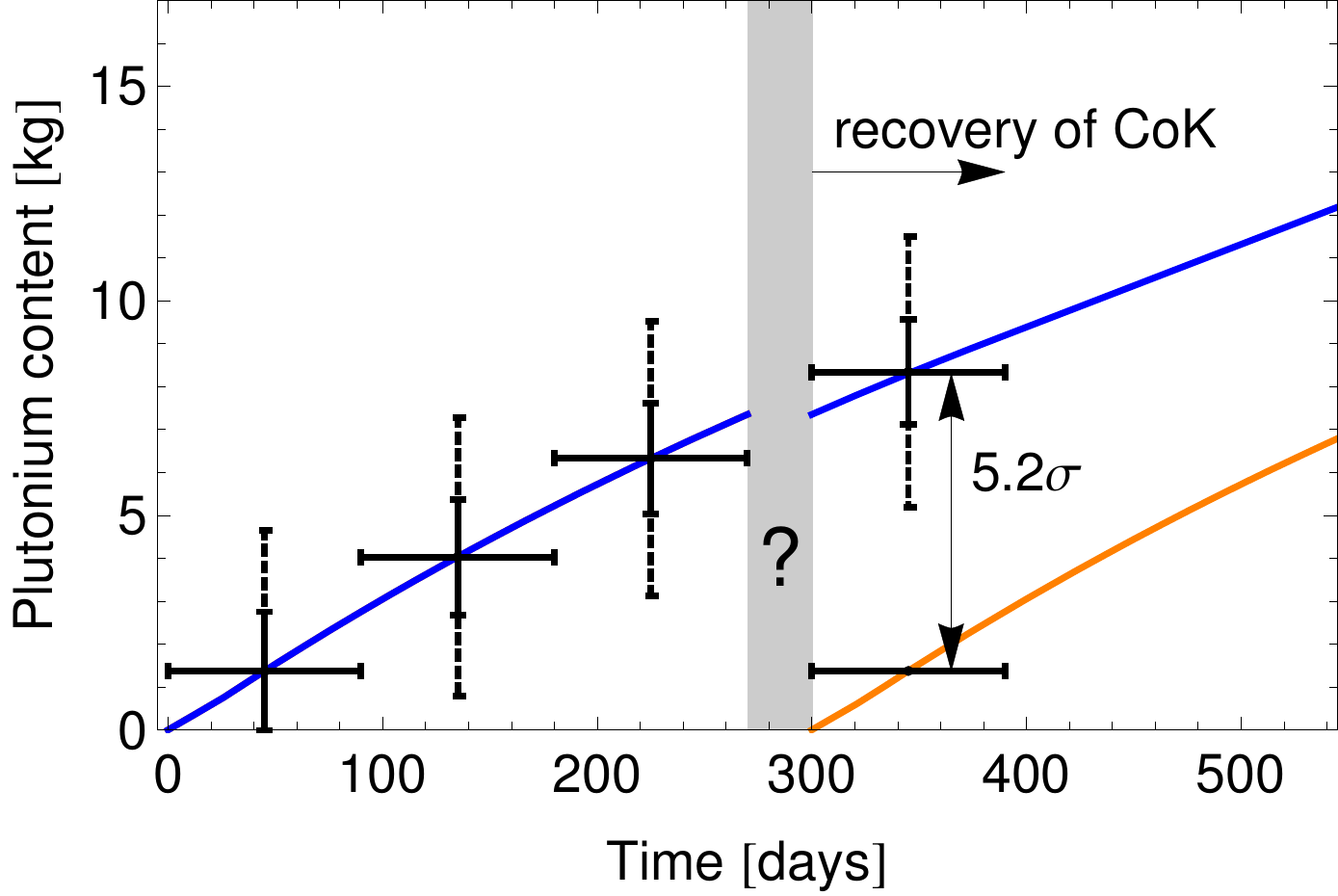}
    \caption{Shown is the 1$\,\sigma$ accuracy for the determination
      of the plutonium content of the reactor as a function of time in
      the reactor cycle. The data taking period is 90 days
      each. Dashed error bars indicate the accuracy from a fit to the
      plutonium fission rate $f_\mathrm{Pu}$, whereas the solid error
      bars show the result of a fit constrained by a burn-up
      model. The blue (dark) line indicates operation without
      refueling and the orange (light) line indicates operation with a
      refueling after 270 days. Figure and caption
      from~\cite{Christensen:2014pva}.}
    \label{fig:pu}
\end{figure}

One case put forward is the so-called N$^\mathrm{th}$-month scenario:
The reactor in question is a heavy-water moderated, natural uranium
fueled 40\,MW$_\mathrm{th}$ reactor, which produces about 10\,kg of
weapons-grade plutonium per full power equivalent year. Assume the
reactor is running at nominal power and that there is full safeguards
access for N-1 months. In the N$^\mathrm{th}$ month, there is a
reactor shutdown followed by a lapse in safeguards access. In month
N+1 reactor operation and safeguards access resume, {\it i.e.} the
inspectors are confronted with a closed reactor core and a running
reactor. Furthermore, if we take $N=10$, then the core just prior to shutdown would
contain 8\,kg weapons-grade plutonium. This is a specific example
for a loss of continuity of knowledge (CoK) incident. Loss of CoK
incidents have been reported and in particular seem to occur in states
which are new to or reentering into the safeguards
regime. Conventional means of safeguards are largely based on item
accountancy and very few actual measurements are ever performed, so
CoK is one of the central pillars. Experience shows that recovery of
CoK in a reactor setting is very difficult, and would be expensive and
highly intrusive, see {\it e.g.}~\cite{Christensen:2013eza}. In
Fig.~\ref{fig:pu} the plutonium mass sensitivity obtained by a
neutrino measurement for the N$^\mathrm{th}$-month scenario is shown.

A 90-day post-shutdown measurement provides a plutonium inventory with
an accuracy of 1.2\,kg or the question of whether the core has been
swapped can be answered with 90\% confidence within 7 days. This
example is based on a 5\,ton detector at 20\,m standoff. It is
important to note, that despite Fig.~\ref{fig:pu} showing data for all
4 measurement periods, the conclusion about the core state really is
obtained in each 90 day period independently of any other 90 day
period. In this scenario, neutrino measurements allow restoration of
the CoK in a short period of time and in an entirely non-intrusive
manner.

In the above example the assumption was that the reactor would be
running at nominal power, but also in the case of the reactor
remaining shut down, there are usable neutrino signatures. These
residual signatures arise from 4 fission fragment nuclides which have
half-lives between 100 days to 28 years. As a result, a reactor core
emits neutrinos even after shutdown. For a time after shutdown between
30 and 90 days, there are 1-2 events per day stemming from the
afterglow. Detection of such a low event rate requires a detector with exceptional background suppression, but given such a detector these events could be used to infer the
presence of an irradiated core with a certain minimum burnup.

For the same reactor and detector combinations, a different fueling
scheme was examined. Assume this reactor, at the same power, was
converted to run on 3.5\% enriched uranium fuel using a light water
moderator~\cite{Willig:2012}. Such a scheme would greatly reduce
plutonium production and extend the fuel cycle. The key to the
neutrino measurement in this case is that the fission rates $f_I$
change significantly faster in a natural uranium fueled reactor than
they do in an enriched core. A measurement of those fission rate
changes, called differential burnup analysis, allows to distinguish
the two fueling schemes within about 180
days~\cite{Christensen:2014pva}.

Burnup also can be determined through a continuous neutrino
measurement of reactor power. The evolution of the total count rate
distinguishes different fuel loadings in a light water reactor: in a
LEU core the rate is expected to decline with time, whereas in a mixed
oxide (MOX) core the rate increase or stays nearly constant. The
rate-based approach has been studied in~\cite{Erickson:2016sdm}
based on highly detailed reactor physics simulations for various MOX
fueling schemes.  A spectral neutrino measurement allows determination
of the fission rates $f_I$ and thus direct confirmation of the
isotopic composition and changes thereof which are expected for a
certain burnup~\cite{Jaffke:2016xdt}. The corollary to those studies
is, that neutrino monitoring can distinguish MOX from LEU and mixed
cores and provide an indication whether reactor-grade or weapons-grade
plutonium is put into the reactor. Neutrino measurements also can
provide assurance that disposition goals in terms of total burnup and
isotopic degradation of weapons-grade plutonium have been met.

Disposition of plutonium in fast breeder reactors has been proposed
and in a broader context, there are fuel cycles, like a thorium-based
one, where fast breeders are an integral part. A breeder reactor is a type of reactor which produces more fissile material than it consumes
and typically is based on the use of fast neutrons. Breeder reactors use driver fuel to generate neutrons and breeding blankets made of fertile material, {\it e.g.} natural uranium or thorium. Due to their use of fast neutrons they can use pure or nearly pure plutonium as driver fuel, whereas in a thermal reactor only relatively limited amounts of plutonium can be added to the uranium fuel, so-called mixed oxide (MOX) fuel. A breeder reactor ran without a blanket of fertile material would be a net user of fissile material and if the driver fuel were made of plutonium, significant quantities of plutonium can be consumed and thus destroyed.
Safeguarding breeder
reactors is complicated by the variable breeding ratio resulting from
the presence/absence of a breeding blanket of fertile
material. Assessing the presence of a blanket is hard because there
are relatively few fissions that occur in the blanket, yet, at the
same time it is placed right next to a vigorously fissioning
core. Effectively, the core fissions drown out any radiation
signatures from the blanket. However, there is a unique neutrino
signature from breeding:
\begin{equation}
{^{238}}\mathrm{U}+n\longrightarrow^{239}\mathrm{U}\stackrel{\beta^-}{\longrightarrow}{^{239}}\mathrm{Np}\stackrel{\beta^-}{\longrightarrow}{^{239}}\mathrm{Pu}\,,
\end{equation}
where the two beta decays have short half-lives of 24\,m and 2.4\,d
and endpoint energies of 1.26\,MeV and 0.72\,MeV,
respectively. Similar signatures exist in a thorium-based fuel
cycle. The resulting antineutrinos are below IBD threshold and hence
invisible to the usual neutrino detectors. It may be possible to
detect them in CE$\nu$NS detectors. A detailed study has been
performed~\cite{Cogswell2016} and the authors found that detectors of
moderate size, several tens of kilograms, could reliably detect the
presence of a breeding blanket at a standoff of 25\,m.

\section{Applications to undeclared reactors: reactor discovery and exclusion}
\label{sec:far}

\subsection{Existing approaches}

Historically, there are numerous cases of reactor construction and
operation being discovered by
intelligence-gathering~\cite{Richelson}. Technological approaches to
discovery or exclusion of reactors have been more
limited. Technological methods that may be useful for remote
monitoring and discovery of reactors include thermal and visible
wavelength satellite or aerial surveillance, and monitoring of xenon,
krypton and other radio-nuclides in the atmosphere far from their point
of origin.

Roughly speaking, a reactor fissions a kilogram of material per GW-day
of heat produced. The heat generated by fission can be rejected into
the air via cooling towers or into a lake, river, or the ocean via
cooling water. These thermal signatures can in principle be detected
from space or airborne thermal-infrared cameras
\cite{Hafemeister1989}, or in the winter, by surface ice melting
downstream from a reactor cooling water outlet. Satellite surveillance
can observe construction activities, and in the case of thermal
imagery, it can provide a rough estimate of power output for some
reactor designs. Disadvantages of this approach are the need for
cueing information, that is extraneous information sources that enable the satellite surveillance to focus the search on a specific area due to its limited field of view, the dependence
on weather, the qualitative nature of the power estimates, and
susceptibility to masking or dissipation of the thermal signature.
 
Noble gases and other radioactive gases from fission are created in
operating reactors.  These can escape through cracks in the outer
layers of fuel rods, and they may ultimately be released to the
atmosphere. The detectability of noble gases released from reactors
depends on the integrity of the fuel and cladding, pathways within the
reactor complex to the atmosphere, and the weather conditions along
the path from the reactor to the radio-nuclide detection
apparatus~\cite{Saey}. This approach to reactor discovery can also
suffer from confounding signals arising from radio-nuclide release from
other nuclear facilities, such as reprocessing plants or radioisotope
production facilities.

Given the relatively limited set of tools available for remote
reactor discovery, exclusion and monitoring, antineutrino-based
methods offer unique features that may be of use in current or future
monitoring regimes.

\subsection{Neutrino-based approaches}

Neutrino-based techniques offer significant advantages: persistence;
the ability to detect or exclude reactor activity in a wide
geographical region without external cueing information; insensitivity
to weather, shielding and other environmental factors; the potential
to place constraints on, or directly measure, the operational status
and total thermal power of the reactor, and thereby estimate the
maximum possible rate of plutonium production in the discovered
reactor.

As standoff distances increase from the near-field regime, the event
rate that can be practically achieved drops, even in large detectors,
from tens or hundreds of events per day to a few events per day, week
or month. Timely direct measurement of fissile content becomes
difficult or impossible, simply due to the small event samples
obtainable in reasonable integration times. Still, it may be possible
to discover, or exclude the existence of, undeclared reactors in
regions surrounding the detector location. In addition, constraints
can be placed on the total power output of a known reactor, or a set
of known reactors, over periods of months, providing an upper bound on
fissile material production. If backgrounds are sufficiently well
understood through simulation and calibration, the existence of an
undeclared reactor can in principle be discovered by looking for a
signal above the known background. If backgrounds must be measured in
place, then only a sufficiently large change in the reactor power can
be observed, manifested as a deviation from a stable background.

Prediction of backgrounds is a significant challenge for these types of experients.  Ambient radioactivity levels  from the detection medium, detector materials and surrounding rock must be measured and incorporated into simulations. As a result, screening campaigns for all construction materials are a common practice for underground particle detectors.  Modeling is more complex for  muogenic backgrounds, including neutrons and long-lived radionuclides. A widely used model for muogenic neutron backgrounds is that of Mei and Hime~\cite{PhysRevD.73.053004}, while muon tranport codes such as MUSIC and MUSUN ~\cite{KUDRYAVTSEV2009339} are used to propagate muons to great depths underground and study angular dependence.

Aside from questions of modeling backgrounds, there are several limitations on neutrino-based approaches: the smallness
of the IBD cross-section; backgrounds of real antineutrinos from the
hundreds of existing civilian power reactors worldwide; and persistence of
cosmic-ray induced backgrounds, which for large detectors can only be
reduced by underground deployment.

We use a 50\,MW$_\mathrm{th}$ reactor as a 'standard candle', this power being roughly typical of the scale of  plutonium production reactors. Excluding the presence of such a reactor  within one year with 95\% confidence at 
1,000\,km standoff requires a 335 kiloton fiducial mass water-based
detector. This mass estimate assumes a 100\% efficient detector above an antineutrino energy threshold of  3.26\,MeV (imposed to remove geoantineutrino backgrounds, as described below), no observed events, and a Poisson-distributed background consistent with zero. With these assumptions,  the 335 kiloton detector would have been  95\% likely to have observed greater than zero events with 3 signal events expected on average. Clearly, the smallness of the IBD cross section is a challenge.

Constraints imposed by other backgrounds further increase the detector size
or dwell time. In order of increasing standoff one needs to deal with
different types of background. Up to 20\,km the dominant backgrounds
are accidentals from local radioactivity, fast neutrons, and
long-lived muogenic radio-nuclides. These can be controlled by locating
the detector underground and by careful material selection. Additional
research is needed to determine the degree to which these backgrounds
can be suppressed in 100 kiloton and larger detectors and studies of
achievable sensitivities have been performed~\cite{SNIF}.

At larger standoffs, geo-neutrinos stemming from uranium and thorium
decays in the earth~\cite{Krauss:1983zn,BELLINI20131} become
non-negligible. Since their energies do not exceed 3.26\,MeV an
energy cut on the reconstructed positron spectrum can remove this
background, though in many detectors, upward fluctuations of the apparent reconstructed energy can contaminate the signal region.

At standoffs of hundred kilometer or more, reactor antineutrino
backgrounds become a limiting factor. These backgrounds are the
greatest concern in monitoring contexts, since they cannot be removed
except by reconstructing the direction of the incident antineutrino,
which is challenging to accomplish for IBD events. Less well measured
but potentially also important are IBD-like events induced by high
energy atmospheric neutrinos and antineutrinos, including both charged
and neutral current channels on oxygen~\cite{PhysRevLett.76.2629}. For
the largest detectors contemplated in this article, at the megaton
scale, the as-yet-unmeasured but long-predicted diffuse supernova
antineutrinos may become a limiting
background~\cite{doi:10.1146/annurev.nucl.010909.083331}.

\begin{table}[t]
\begin{tabular}{lcccccc}

Distance [km] & 10   & 20  & 50 & 100  & 200  \\ 
\hline
\hline
Low background    & 1$\times$0.08 & 1$\times$0.4 & 10$\times$1     & 100$\times$1  & 1000$\times$0.8         \\ \hline 
Medium backgr. & 1$\times$0.1  & 1$\times$0.7 & 100$\times$0.7  & 1000$\times$1 &                 \\ \hline 
High background    & 1$\times$0.3  & 5$\times$1   & 1000$\times$0.9 &        &  \\ \hline              
\end{tabular}
\caption{For three representative reactor antineutrino background
  levels, this table shows the detector fiducial mass in kilotons and
  dwell time in years required to achieve $3~\sigma$ sensitivity to
  the presence of a 50 MW$_\mathrm{th}$ reactor. The three background
  categories correspond to the actual reactor and geo-neutrino
  backgrounds at the existing Andes, Baksan and Frejus underground
  laboratories~\cite{Barna:2015}, with 170, 2,080 and 28,000
  background events per year and 100 kiloton detector mass. The data
  is formatted as mass [kt]$\times$dwell time [years]. Blank cells
  indicate that the dwell time is greater than 1 year, or the detector
  mass is greater than 1 megaton. Neutrino oscillations are accounted
  for and an energy cut to largely remove geo-neutrinos is applied.}
 \label{tab:sizevstandoff}
\end{table}
The summed background contributions from all of the world's reactors
at any point on Earth can be estimated to a precision of about $5\%$
\cite{Usman:2015yda,Barna:2015}. This integrated background
contribution varies by a factor of about 30 from the Northern to
Southern hemisphere, ranging from a high of 2000 to a low of $\sim$65
events per 100 kiloton of water per year~\cite{SNIF}. This background
is irreducible, unless event-by-event measurements of the neutrino
direction become possible. Therefore, the limit in sensitivity is set
by the global reactor neutrino background, where we distinguish
regions with low, medium and high reactor neutrino background,
respectively, as shown in Table~\ref{tab:sizevstandoff}. The conclusion
from this simple exercise is that standoff distances beyond 200\,km
will require event-by-event measurements of the neutrino
direction~\cite{Jocher:2013gta}.

\subsection{Technology options}
\label{sec:techopts}
 
Water Cerenkov and scintillation detectors are the only viable target
media for the construction of large-scale (kiloton and above)
antineutrino detectors implied by Table~\ref{tab:sizevstandoff}. Within tens of kilometers, few-kiloton
detectors suffice to achieve basic monitoring goals, {\it e.g.} KamLAND~\cite{Eguchi:2002dm} and JUNO~\cite{An:2015jdp}, and these could be based on liquid scintillator. To build the larger, 100 kiloton or megaton size, detectors for use in the far-field, water-based technologies appear promising. The 50 kiloton Super-Kamiokande water Cerenkov detector has already demonstrated
sensitivity to MeV-scale (solar)
neutrinos~\cite{PhysRevLett.112.091805}. However, in pure water
detectors, neutrino and antineutrino are indistinguishable. This greatly complicates the detection of antineutrinos, because the neutrino/antineutrino signal consists only of a single flash of light induced by the neutrino or antineutrino. For that signal, backgrounds consist of the full gamut of sources that can induce MeV-scale single events, including gamma-rays from radioactive contaminants in the target medium and detector materials, cosmogenic muons and neutrons, and muogenic radionuclides. Solar and other neutrinos are also of course a background to antineutrinos in such detectors. Conversely, if the neutron from IBD interactions (see Eq.
~\ref{eq:IBD})  can be tagged efficiently, the presence of this signal in close time coincidence with that induced by the positron permits suppression of backgrounds by 3 orders of magnitude or more compared to a search for a single MeV-scale energy deposition. 

To break the degeneracy of antineutrino and neutrino, and permit efficient and unambiguous detection of
MeV-scale antineutrinos, researchers have proposed to add gadolinium
to water~\cite{Bernstein2001,Beacom:2003nk}, at roughly the
part per thousand level by weight. Gadolinium, an efficient
neutron-capture element, greatly improves the efficiency for
detection of the final state neutron in the IBD process.

A 200-ton engineering demonstration of gadolinium-doped water
technology has been achieved by the EGADS group~\cite{EGADS}. The
experiment demonstrated the compatibility of standard materials with
gadolinium-doped water, and showed that the effective attenuation
length of Cerenkov light in gadolinium-doped water remained high, a
key consideration for the construction of large-scale detectors. In
part based on this research, the Super-Kamiokande collaboration
announced~\cite{SKGDnews} that it would add gadolinium to the
detector, primarily in an effort to detect diffuse supernova
antineutrinos.

In 2018, the dedicated WATCHMAN experiment was launched~\cite{WATCHMANnews} to investigate the viability and
scalability of gadolinium-doped water as a tool for reactor
antineutrino detection in nonproliferation contexts. It will be
constructed in the Boulby mine in Northern England, and will measure
neutrinos emitted by the Hartlepool nuclear reactor complex, 25
kilometers distant. 

In order to breach the 200 kilometer limit for remote sensitivity
implied by Table~\ref{tab:sizevstandoff}, directional reconstruction
methods on an event-by-event basis will be needed for reactor
antineutrinos. In the IBD reaction the momentum of the neutrino is
carried by the neutron and hence the neutron momentum would need to be
reconstructed, a daunting task in a megaton-scale detector. In the
neutrino-electron scattering reaction, the scattered electron carries
the momentum of the neutrino, but the expected event rate per unit
mass for hydrogenous targets is approximately 5 times lower than for
IBD~\cite{Dye:2017scn}. 

In spite of the difficulties, the high value of directional reconstruction for background suppression motivates continued investigations in this area. Examples of directional concepts for IBD and neutrino-electron scattering respectively are found in ~\cite{PhysRevLett.114.071802}  and ~\cite{Hellfeld:2015xym}. 


\section{Applications to spent fuel and reprocessing waste: discovery and monitoring}
\label{sec:waste}

\subsection{Existing approaches}

At present, compliance with safeguards agreements is based on
observations made before a storage cask or underground repository is
closed and relies upon the integrity of seals and remotely monitored
cameras to verify that these closed volumes were not opened between
inspector visits. However, seals can be opened and closed without
detection~\cite{Johnston2002} and cameras can be unplugged or blocked,
intentionally or inadvertently. It therefore would be desirable to
verify that the situation inside a sealed container or repository is
as expected without having to open it.

It is challenging to verify the plutonium content of SNF using
nondestructive measurements. IAEA SNF safeguards therefore employ
radiation measurements to confirm that specific characteristics
(termed attributes) of the fuel are consistent with the declared
\ph{initial enrichment, cooling time and burnup}. These radiation
measurements \ph{are typically confined to take place during wet
  storage in a fuel pond and are performed using a} combination of
gamma spectroscopy, gross neutron counting, neutron coincidence
counting, and Cerenkov imaging\ph{, where the latter is the most
  commonly used. In principle, this combination allows} to confirm
gamma and neutron emissions expected from characteristic nuclides and
Cerenkov light indicating that all individual fuel rods in the
assembly are present\ph{, see also Sec.~\ref{sec:reactor_existing}.}

\subsection{Neutrino-based approaches}
\ph{Dry storage of some form is the final destination for almost all
  SNF.  The bulk of SNF is currently in wet storage, but in
  the aftermath of the Fukushima Daichi nuclear accident the
  associated safety ramifications became all too
  obvious~\cite{NAP21874}. These safety concerns combined with
  eventual decommissioning of nuclear power plants will lead to a
  significant increase of the amount and fraction of all SNF in dry
  storage, see for instance~\cite{GAOSNF}.} For the verification of
SNF in dry-cask storage facilities, neutrino monitoring could be an
option if the cost were affordable within the IAEA budget. SNF and
reprocessing waste (RW) are intensely radioactive and the bulk of
nuclear decays occurs via beta decay, thus, both constitute neutrino
sources. For neutrino detection based on IBD, however, only neutrinos
above the IBD threshold of 1.8\,MeV are visible. Sargent's rule states
that beta decay rates are proportional to $Q^5$, where $Q$ is the
endpoint energy in the neutrino spectrum; thus beta emitters with an
endpoint above the IBD threshold tend to be very short-lived. One year
after discharge from reactor, all detectable neutrinos stem from only
3~pairs of nuclides: $^{90}$Sr/Y, $^{144}$Ce/Pr and $^{106}$Ru/Rh. The
reason they appear in pairs is related to Sargent's rule: the first
decay in the pair is a low-energy, and hence relatively long-lived
decay, whereas the second decay is of higher energy and therefore
short-lived. For source material older than a few years, only the
$^{90}$Sr/Y decay chain, with a half-life of about 29~years, is
relevant. This also implies that for any SNF/RW produced to date, only
about 2.6~half-lives have elapsed, and this emission is still at 16\%
of its original value. Fortunately, $^{90}$Sr has a high cumulative
fission yield\footnote{Cumulative fission yield is the sum of the number of atoms per fission produced directly by the fission and those arising from decays of other fission products.} of 1--5\%. In reprocessing, $^{90}$Sr will end up in the
waste stream and thus RW is a significant neutrino source for long
periods of time.

In most countries, the bulk of SNF produced in commercial nuclear
power plants eventually ends up in dry storage casks. The rate of
neutrino events per ton of fiducial detector mass and per metric ton
of uranium (MTU) of source mass is, assuming a burnup of
45\,GW\,d\,$\text{MTU}^{-1}$~\cite{Brdar:2016swo}
\begin{align}
  N_\nu = 5.17\;\text{yr}^{-1} \, \text{ton}^{-1} \, \text{MTU}^{-1}
            \times ( \text{10\,\text{m}}/L )^2 \,,
\end{align}
where $L$ is the distance between the source and the detector (both
treated as point-like). Typically, these storage facilities are close
to an operating nuclear reactor complex and thus there will an
irreducible background of neutrinos coming from the reactor. This size
of this background can be accurately measured in the same neutrino
detector used for the SNF signal; due to the high energy of the
reactor neutrinos as compared to the SNF neutrinos the two components
can be disentangled and only the statistical uncertainty from
background subtraction remains. In~\cite{Brdar:2016swo}, a real
existing dry storage facility is taken as an example and it is found
that a change of inventory by as little as 3\% can be detected with
exposures in the range of 20--80 ton years at a stand-off of up to
50\,m. In this analysis the assumption is made that cosmogenic and
other non-neutrino backgrounds can be reduced to negligible levels.

Eventually, most nations plan to store SNF in long-term geological
repositories. Given the large amount of SNF at such a site,
$10^4-10^5$~MTU, the resulting neutrino signal will be large, tens of
events per year and ton at kilometer scale standoff. In particular,
after closure of the repository, neutrinos will be the only detectable
radiation signature. Following the analysis in~\cite{Brdar:2016swo},
however, the total large amount of SNF makes it difficult to be
sensitive to quantities of interest either in the context of
non-proliferation or safety of the repository: even the loss of 1 cask
with a few MTU, in either case, would be significant, but this is far
less than 1\% of the inventory. Effectively the remaining 99.x\% of
SNF blinds the neutrino detector. This situation would improve, if
directional neutrino detection in large detectors, 100s or 1000s of
tons, became available, which potentially could be achieved by liquid
argon time projection chambers, as discussed in 
Sec.~\ref{sec:techopts}.

Industrial-scale reprocessing results in significant quantities of
liquid, highly radioactive wastes. Historically, for the nuclear
weapons programs of the US and USSR, these wastes have been
stored in underground tank farms and their corrosion presents a major
problem due to the risk of ground water
contamination~\cite{Jaraysi:2006,Rockhold:2012}. Given that $^{90}$Sr
is extracted into the aqueous phase in the PUREX process, these RW
tanks also contain large quantities of $^{90}$Sr and thus are the
source of detectable neutrino emissions. In~\cite{Brdar:2016swo}
a study of a tank farm based on an existing site~\cite{Jaraysi:2006},
shows that a 80-year-ton exposure can measure the $^{90}$Sr content of
a given tank at the 20\% level. Equivalently, for a known quantity of
reprocessed fuel this allows an age determination of in the range of
44-54 years for a true age of 50 years. This capability could be
useful in clarifying the history of a plutonium-based weapons program.

In the previous example, the location of the RW was known but the
quantity was not. The logical extension is the case where also the
location is not known precisely. This situation could arise naturally
when undeclared reprocessing is suspected and the goal is obtain a
rough estimate of the possibly extracted amount of plutonium. Such a
scenario was encountered by the IAEA in 1992 in dealing with North
Korea: isotopic analysis of samples taken during inspection indicated
three reprocessing campaigns, whereas the initial declaration stated a
single reprocessing campaign. The use of a neutrino detector
specifically for this case has been subject of a detailed
study~\cite{Christensen:2013eza}: a complete reactor core of the
5$\,\mathrm{MW}_\mathrm{e}$ reactor corresponds to about 8\,kg of
plutonium if fully reprocessed. The resulting RW can be detected at a
standoff of 25\,m with an exposure as little as 1--2 ton years and at
a standoff of 100\,m with an exposure in the 50--200 ton year
range. The large increase in required exposure is due the background
from the operating reactor nearby; otherwise, required exposure simply
would increase as the square of the standoff.

\section{Applications to nuclear explosions: fission confirmation and yield estimation}
\label{sec:weapons}

\subsection{Existing approaches}

The CTBT verification regime relies in part on the International
Monitoring System (IMS), a global network of facilities to detect
nuclear explosions. Seismic~\cite{Kværna2013},
hydro-acoustic~\cite{Lawrence1999}, infrasound~\cite{Green2010}, and
radio-nuclide verification~\cite{Schoeppner2017} technologies comprise
the IMS and are distributed across 337 stations and laboratories to
monitor for nuclear explosions conducted on
Earth~\cite{IMSFacilityMap}. Currently, the most sensitive means for
detecting underground nuclear explosions are seismic, which can detect
and identify explosions down to or below a yield of about 1 kiloton
worldwide. At low yields, if radioactive gases do not leak out in
detectable quantities, it is theoretically possible that an explosion
could be claimed to be conventional (although mining explosions are
typically ripple-fired blasts, which are seismically distinguishable
from a nuclear explosion). A nuclear explosion under the ocean would
be detectable via hydroacoustic waves and in the atmosphere by the
characteristic double pulse of light and radioactive fallout. In
space, detection satellites monitor for a pulse of X-rays
\cite{NAP2012}.

\subsection{Neutrino-based approaches}

For a WATCHMAN-sized Gd-doped water detector, $10^3\,\mathrm{m}^3$
fiducial volume, detection of antineutrinos in coincidence with
seismic events could in theory provide unambiguous signatures of a
kiloton fission explosion out to a few km and a 250 kiloton explosion
out to a few tens of km. The largest proposed detector, with a
fiducial volume of $\sim$ 200,000\,m$^3$ could detect a 1 kiloton
fission explosion at a distance of about 20\,km~\cite{Carr2017}.  With
fiducial volumes on the order of $10^8\,\mathrm{m}^3$ detectors of
this type would be able to detect 1 kiloton fission explosions at a
distance of 1000\,km or a 100 kiloton fission explosion at a distance
of 10,000\,km, providing global coverage.

\section{Summary \& outlook}

The pursuit of practical roles for neutrinos, especially in nuclear
security, goes back at least 40 years. In those four decades, our
understanding of fundamental neutrino properties has improved
considerably, and neutrino emissions from fission sources have been
more precisely characterized. Multiple detection channels have come
into use, and the IBD channel has become a workhorse for fundamental
science. As we have highlighted, neutrinos were first detected at a
reactor producing plutonium for nuclear weapons. In this sense, the
science of neutrinos and the wider uses of nuclear fission technology
have long shared a link.

Any successful application of neutrinos will reconcile their unique
advantage as a fission signature -- the ability to pass through large
amounts of matter -- with the flip side of that property, the
difficulty of identifying these particles in significant numbers in a
realistic detector. This central constraint favors applications in
which the flux of neutrinos is high. Of the three fission sources
considered here, operating reactors have the highest time-averaged
flux on timescales relevant for security problems, hours to months, at
distances reasonable for observation, several meters to hundreds of
kilometers.

For this reason, reactors are the most promising target for neutrino
applications in the near term. As we have outlined, neutrinos may be
useful for two different regimes of reactor monitoring. The first case
is near-field monitoring, $\lesssim1$ km standoff, of known
reactors. In near-field scenarios, few-ton-scale scintillator
detectors with linear dimensions of several meters can detect on/off
transitions, track power levels, meet IAEA standards for spotting
plutonium diversion, and meaningfully track plutonium
disposition. Detector technologies providing the requisite energy
resolution and background rejection have been recently
demonstrated. With modest further investment, these technologies could
be deployed as a real-time, less invasive complement to existing
reactor verification techniques.

A second and more ambitious application for reactor neutrinos is
discovery of hidden, undeclared reactors. This capacity would be most
valuable when the sensitive range of the detector covers distances of
several hundred kilometers or more, extending over wide territories
and possibly national boundaries. That aspiration calls for detectors
as large as the multi-megaton scale with 100\,m or larger in linear
dimensions. While the engineering challenges and costs of
megaton-scale detectors are formidable, systems on this scale are
under active development for basic science. However, the background
stemming from known civilian nuclear reactors presents a major obstacle
and only event-by-event measurement of the neutrino direction can
overcome this limitation. On the other hand, for the distance range
from 10's to 100's of kilometers, the key enabling technologies for
suitably large detectors are well developed: in the next decade, the
WATCHMAN program expects to demonstrate reactor discovery capabilities
in a 1\,kiloton fiducial mass detector at a distance of 25\,km~\cite{Askins:2015bmb}.

While the other stages in the nuclear fuel cycle offer opportunities
for neutrino monitoring, they present considerably more challenging
detection problems than operating reactors. The emission rates and
energies of neutrinos emitted from SNF and reprocessing
waste are lower than from reactors. Still, ton-scale scintillator
detectors offer rare capabilities for verifying the contents of sealed
spent fuel casks and identifying well-concealed reprocessing
waste. The burst of neutrinos following an underground nuclear weapon
test could help formally identify its fission nature when combined
with seismic data. However, even megaton-scale detectors could surveil
only a limited geographic region and would minimally enhance the
strong forensic power of the existing explosion monitoring network.

This review focused on mature technologies, namely detectors for IBD,
which has now been observed over five million times in basic science
experiments at nuclear reactors. Technologies continuing to emerge
from basic science, such as detectors for CE$\nu$NS, may eventually
create new application options. CE$\nu$NS offers the possibility of
detecting neutrino from breeding reactions, which are below IBD
threshold, and may allow for smaller active detector masses. CE$\nu$NS
has been observed for the first time in 2017~\cite{Akimov:2017ade}
with neutrinos from a spallation neutron source, yet no confirmed
detection of reactor neutrino via this reaction exists. The first
definitive measurement of CE$\nu$NS from the reactor neutrino signal
will likely first be accomplished with ionization-based
detectors. However, such detectors suffer an impractical limit on
their minimum size, essentially imposed by the relatively large amount
of energy, $10-20\,\mathrm{eV}$, needed to create a single ionization
event.  To realize practical detectors that are smaller than IBD
detectors at a given standoff, very low-threshold, ({\it e.g.}
phonon-sensitive) CE$\nu$NS detectors will need to be developed, then
scaled to useful sizes. Directionality and spectroscopy via the CE$\nu$N
channel are even more difficult to achieve.  As a result,
CE$\nu$NS-based approaches are unlikely to compete with IBD-based
monitoring for a decade or longer. Note, in the case of IBD it took
more than 60 years from a first detection to detectors which are
capable of a safeguards mission.

Over several decades, physicists have conceived many ideas for using
fission neutrinos in nuclear security. Some ideas remain in the realm
of pen and paper, constrained by basic physical and practical
considerations. For other concepts, demonstrated technology is
catching up with real opportunities. The unique safeguards
capabilities provided by near-field monitors, in particular the
ability to recover lost continuity of knowledge, make a first
application more likely in cases where there is a lack of a
well-established history of safeguards and mutual trust. This seems to
favor applications within the verification provisions of bi- or
multi-lateral agreements between nations, instead of a regular
safeguards agreement between a nation and the IAEA. In this context,
also cost would be much less of a concern.  For near-field reactor
monitoring in particular, technology now exists to support the first
on-the-ground applications.

\section*{Acknowledgments}

The authors wish to acknowledge extensive discussions with Rachel Carr
and Frank von Hippel that took place during the preparation of this
review, and Viascheslav Li for calculations of oscillated spectra and rates for reactors at large standoff. 

LLNL-JRNL-784940. This work performed under the auspices of the US
Department of Energy by Lawrence Livermore National Laboratory under
Contract DE-AC52-07NA27344. This material is based upon work supported
in part by the Department of Energy National Nuclear Security
Administration Office of Defense Nuclear Nonproliferation R\&D through
the Nuclear Science and Security Consortium under Award Number
DE-NA0003180, through the Consortium for Verification Technology
under Award Number DE-NA0002534, and through the consortium for Monitoring, Technology and Verification under Award number DE-NA0003920. PH was supported by the US Department
of Energy Office of Science under award number DE-SC0018327.

\bibliography{nu_review}

\end{document}